\documentclass[]{interact}

\usepackage{epstopdf,multirow}% To incorporate .eps illustrations using PDFLaTeX, etc.
\usepackage[caption=false]{subfig}% Support for small, `sub' figures and tables

\usepackage[numbers,sort&compress]{natbib}% Citation support using natbib.sty
\bibpunct[, ]{[}{]}{,}{n}{,}{,}% Citation support using natbib.sty
\makeatletter% @ becomes a letter
\def\NAT@def@citea{\def\@citea{\NAT@separator}}% Suppress spaces between citations using natbib.sty
\makeatother% @ becomes a symbol again

\theoremstyle{plain}% Theorem-like structures provided by amsthm.sty

\theoremstyle{definition}

\theoremstyle{remark}

\begin{document}

%%%%%LAURA
\def\T{{\footnotesize {^{_{\sf T}}}}} 
\newcommand{\Real}{{\rm I}\negthinspace {\rm R}}
\newcommand{\sthat}{\widehat}
\newcommand{\I}{\mathbb{I}}

\newcommand{\E}{\mbox{\rm E}}
%%%%%%%%%%%%%

\title{Robust confidence distributions from proper scoring rules}

\author{
\name{Erlis Ruli\textsuperscript{a}, Laura Ventura\textsuperscript{a} and Monica Musio\textsuperscript{b}}\affil{\textsuperscript{a} University of Padova, Italy, {\tt ruli@stat.unipd.it, ventura@stat.unipd.it}; \\
 \textsuperscript{b} University of Cagliari, Italy, {\tt mmusio@unica.it}}
}

\maketitle

\begin{abstract}
A confidence distribution is a distribution for a parameter of interest based on a parametric statistical model. As such, it serves the same purpose for frequentist statisticians as a posterior distribution for Bayesians, since it allows to reach point estimates, to assess their precision, to set up tests along with measures of evidence, to derive confidence intervals, comparing the parameter of interest with other parameters from other studies, etc. A general recipe for deriving confidence distributions is based on classical pivotal quantities and their exact or approximate distributions. 

However, in the presence of  model misspecifications or outlying values in the observed data, classical pivotal quantities, and thus confidence distributions, may be innacurate. The aim of this paper is to discuss the derivation and application of robust confidence distributions. In particular, we discuss a general approach based on the Tsallis scoring rule in order to compute a robust confidence distribution. Examples and simulation results are discussed for some problems often encountered in practice, such as  the two-sample heteroschedastic comparison, the receiver operating characteristic curves and regression models.
\end{abstract}


%%%%%%%%%%%%%%%%%%%%%%%%%%%%%%%%%%%%%%%%%%%

\section{Introduction}

Suppose data are analysed via some parametric model, and that $\psi$ is a parameter of interest, such as a location parameter, the difference between two means, the Area Under the Roc Curve (AUC), a regression coefficient, etc. For conducting inference on $\psi$, statisticians have many methods in their toolboxes, such as reaching point estimates, assessing their precision, setting up tests along with measures of evidence, finding confidence intervals, comparing $\psi$ with other parameters from other studies, etc. All these inference topics may be automatically performed using an unique tool when a frequentist distribution for $\psi$, given the observed data, is available. 

\begin{figure}
\begin{center}
\includegraphics[scale=0.5,height=7cm]{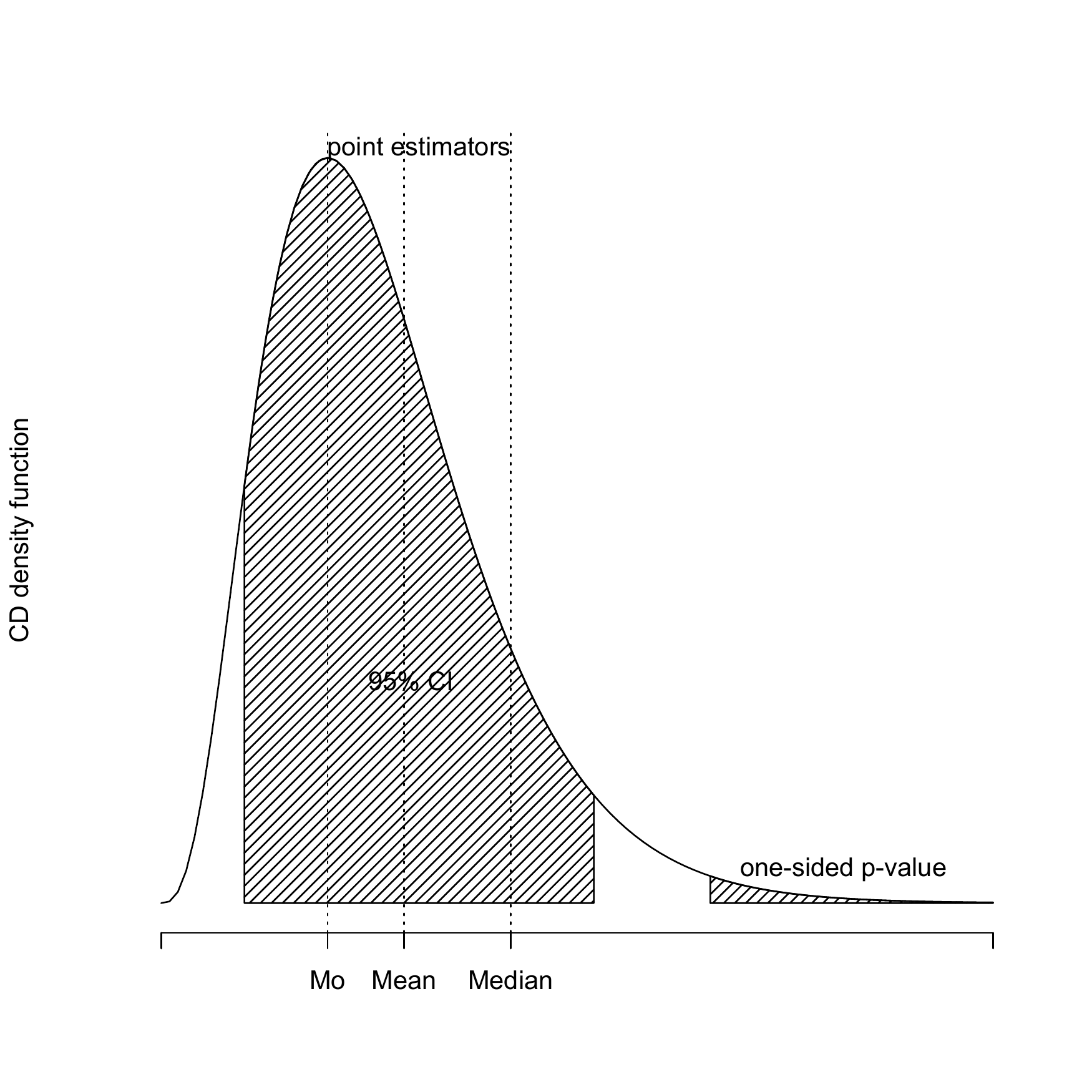}
\vspace{-0.4cm}
\caption{{\small Illustration of making inference on the scalar parameter of interest $\psi$ using a confidence density.}}
\label{fig2}
\end{center}
\end{figure}

A practical approach on how to reach proper frequentist distributions, without priors, is based on confidence distributions (CDs) and confidence curves (CCs); see, among others, Xie and Singh (2013),  Schweder and Hjort (2016), Hjort and Schweder (2018), and references therein. In practice, a confidence curve analysis is much more informative than providing the prototypical 95\% interval or a $p$-value for an associated hypothesis test. The plot in Figure \ref{fig2} gives an illustration on making inference using a CC: point estimators (mode, median and mean), 95\% confidence interval and one-sided $p$-value. 

The standard theory for parametric models evolves around the use of likelihood methods, and this is also partly the case for the theory and applications of CDs and CCs. Typically, to first-order, CD inference may be based on familiar large-sample theory for the maximum likelihood estimator (MLE), the Wald statistic and the likelihood-ratio test. The basic concepts and recipes for CDs and CCs are however not limited to likelihoods per se, and various alternatives may be worked with. For instance, it is well-known that for  model misspecifications or in the presence of deviant values in the observed data, likelihood methods may be innacurate in many applications (see, e.g., Heritier {\em et al.}, 2009 and Farcomeni and Ventura, 2012). To deal with model misspecifications, it may be preferable to base inference on procedures that are more resistant, that is, which specifically take into account the fact that the assumed models used by the analysts are only approximate. In order to produce statistical procedures that are stable with respect to small changes in the data or to small model departures, robust statistical methods can be considered.

In this paper, we focus in particular on robust procedures based on proper scoring rules (SRs). A scoring rule (see, for instance, the overviews by Machete, 2013, and Dawid and Musio, 2014, and references therein) is a special kind of loss function designed to measure the quality of a probability distribution for a random variable, given its observed value. Proper scoring rules supply unbiased estimating equations for any statistical model, which can be chosen to increase robustness or for ease of computation. The Brier score (Brier, 1950), the logarithmic score (Good, 1952), the Tsallis score (Tsallis, 1988), and the Hyv\"arinen score (Hyv\"arinen, 2005) are well-known instances of scoring rules. In particular, when using the logarithmic score, the full likelihood and the composite likelihood (Varin {\em et al.}, 2011) are obtained as special cases of proper scoring rules (see for instance Dawid and Musio, 2014). Frequentist scoring rule inference has been widely discussed, also for robustness (see Ghosh and Basu, 2013, Dawid {\em et al.}, 2016, and references therein), while Bayesian inference based on scoring rules has been considered in Dawid and Musio (2015), Ghosh and Basu (2016) and Giummol\'e {\em et al.} (2019). To our knowledge, the only application of a particular scoring rule in the context of CD inference is illustrated in Hjort and Schweder (2018, Sect.\ 7). However, the associated scoring rule likelihood-ratio type test has not a standard $\chi^2$ distribution, and thus is not an appropriate pivotal quantity. 

The aim of this paper is to discuss a general approach based on the Tsallis scoring rule in order to compute a robust CD. In particular, we explore asymptotic robust SR-CDs obtained by using pivotal quantites from SR inference (see Dawid {\em et al.}, 2016). Examples and simulation results are discussed for three problems which statisticians often encounter in practice, such as the two-sample heteroschedastic $t$-test, the area under the receiver operating characteristic (AUC) curve and regression models. 

The paper unfolds as follows. Sections 2 and 3 review, respectively, some background on CDs and scoring rules. Section 4 discusses the construction of the proposed SR-CDs and the derivation of the corresponding tail area influence function. Examples and simulations studies are presented in Section 5. Finally, concluding remarks can be found in Section 6.

%%%%%%%%%%%%%%%%%%%%%%%%%%%%%%%%%%%%%%%%%

\section{Background on confidence distributions}

Consider a random sample $y=(y_1,\ldots,y_n)$ of size $n$ from  a parametric model with probability density function $f(y;\theta)$, indexed by a $d-$dimensional parameter $\theta$. Write $\theta=(\psi,\lambda)$, where $\psi$ is a scalar parameter for which inference is of interest and $\lambda$ represents the remaining $(d-1)$  nuisance parameters.  

A modern definition of a confidence curve for $\psi$, say $cc(\psi)=cc(\psi,y)$, can be found, among others, in Xie and Singh (2013) and Schweder and Hjort (2016); see also references therein. Write $Y$ for the random outcome of the data generating mechanism. At the true parameter point $\theta_0=(\psi_0,\lambda_0)$, the random variable $cc(\psi_0)=cc(\psi_0,Y)$ should have a uniform distribution on the unit interval. Then
\begin{eqnarray*}
P_{\theta_0} (cc(\psi_0,Y) \leq \alpha ) = \alpha, \quad  \text{for all} \, \, \alpha.
\end{eqnarray*}
Thus confidence intervals can be read off, at each desired level. When $\alpha$ tends to zero the confidence interval tends to a single point, say $\tilde\psi$, the zero-confidence level estimator of $\psi$. 

In regular cases, $cc(\psi)$ is decreasing to the left of $\tilde\psi$ and increasing to the right, in which case the confidence curve $cc(\psi)$ can be uniquely linked to a full confidence distribution $C(\psi)=C(\psi, y)$, via
\begin{eqnarray*}
cc(\psi) = |1-2C(\psi,y)| = \left\{ \begin{array}{ll}
1-2C(\psi,y), & \text{if}  \, \, \psi \leq \tilde\psi \\
2C(\psi,y)-1, & \text{if}  \, \, \psi \geq \tilde\psi.
\end{array} \right.
\end{eqnarray*}
With $C (\psi)$ a CD, $[C^{-1} (0.05), C^{-1} (0.95)]$ becomes an equi-tailed 90\% confidence interval, etc. Also, solving $cc(\psi) = 0.90$ yields two cut-off points for $\psi$, precisely those of a 90\% confidence interval. Correspondingly one may start with a given set of nested confidence intervals, for all levels $\alpha$, and convert these into, precisely, a CD. 

A general recipe to derive a CD is based on pivotal quantities. Suppose $q(\psi;y)$ is a function monotone increasing in $\psi$, with a distribution not depending on the underlying parameter, i.e.  $q(\psi;y)$ is a pivotal quantity. Thus $Q(x) = P_\theta (q(\psi ;Y ) \leq x)$ does not depend on $\theta$, or on $\psi$, which implies that
\begin{eqnarray*}
C(\psi) = Q(q(\psi;y))
\end{eqnarray*}
is a CD. The corresponding confidence curve for $\psi$ is
\begin{eqnarray*}
cc(\psi) =  \frac{\partial Q(q(\psi;y))}{\partial q(\psi;y)} \, \frac{\partial q(\psi;y)}{\partial \psi}.
\end{eqnarray*}

\subsection{Likelihood-based CDs}

In various classical setups for parametric models, there are well-working large-sample approximations for the behaviour of estimators, etc., and these lead to constructions of CDs and CCs. For instance, if an estimator $\hat\psi$ is such that $(\hat\psi -\psi) \, \, \dot\sim \, \, N(0,\tau^2)$ to first-order, and $\hat\tau$ is a consistent estimator for $\tau$, then $(\hat\psi-\psi)/\hat\tau \, \, \dot\sim \, \, N(0, 1)$. Writing 
\begin{eqnarray*}
C(\psi) \, \dot{=} \, \Phi \left(  \frac{ \psi-\hat\psi}{\hat\tau} \right),
\end{eqnarray*}
we have $C(\psi) \, \, \dot\sim \, \, U(0,1)$. Hence such $C(\psi)$ is an asymptotically first-order valid CD, allowing us to write 
\begin{eqnarray*}
\psi|y \, \,  \dot\sim \, \, N(\hat\psi,\hat\tau^2),
\end{eqnarray*}
in the CD sense. In particular, if $\hat\psi$ is the MLE of $\psi$, then the CD is derived from the profile Wald statistic 
\begin{eqnarray}
w_p(\psi) = \frac{\hat\psi-\psi}{\sqrt{j_p(\hat\psi)^{-1}}},
\label{cdwald}
\end{eqnarray}
with $j_p(\psi)$ profile observed information,
and it coincides with the asymptotic first-order Bayesian posterior distribution for $\psi$. 

A recipe that typically works better than (\ref{cdwald}) is the following. Let $\ell(\theta)$ be the log-likelihood function for $\theta$, and let $\ell_p (\psi) = \ell(\psi,\hat\lambda_\psi)$ be the profile log-likelihood for $\psi$, where $\hat\lambda_\psi$ is the MLE for $\lambda$ given $\psi$. The profile  log-likelihood ratio test $W_p(\psi) = 2 (\ell_p(\hat\psi) - \ell_p(\psi))$, under mild regularity conditions, has an asymptotic null $\chi^2_1$ distribution. Hence $ \Gamma_1(W_p(\psi)) \, \, \dot\sim \, \, U(0,1)$, with $\Gamma_1(\cdot)$ denoting the $\chi^2_1$ distribution function, and 
\begin{eqnarray*}
C(\psi) \, \dot{=} \, \Gamma_1(W_p(\psi))
\end{eqnarray*}
is an asymptotic CD. It can reflect asymmetry and also likelihood multimodality in the underlying distributions, unlike the simpler Wald-type confidence distribution. Similarly, the profile likelihood root
\begin{eqnarray*}
r_p (\psi) = \text{sign} (\hat\psi-\psi) \sqrt{2(\ell_p(\hat\psi)-\ell_p(\psi))}
\label{r}
\end{eqnarray*}
can be used to derive a first-order CD, since it has a first-order standard normal null distribution. 
Improved CD inference based on higher-order asymptotics is discussed in Ruli and Ventura (2020).

\section{Background on scoring rules}

It is well-known that for complex models and/or model misspecification, likelihood methods may be innaccurate in many applications. To deal with complex models or model misspecifications, useful surrogate likelihoods are given by proper scoring rules. 

A scoring rule is a loss function which is used to measure the quality of a given probability distribution $Q$ for a random variable $Y$, in view of the result $y$ of $Y$; see Dawid (1986). The function $S(y;Q)$ takes values in $\mathbb{R}$ and its expected value under $P$ will be denoted by $S(P;Q)$. The scoring rule $S$ is called proper relative to the class of distributions $\mathcal{P}$ if  the following inequality is satisfied for all $P,\,Q\in\mathcal{P}$:
\begin{equation*}
\label{eq:1}
S(P;Q)\geq S(P;P).
\end{equation*}
It is strictly proper relative to $\mathcal{P}$ if equation (\ref{eq:1}) is satisfied with equality if and only if $Q = P$. Note that in the following we identify a distribution $Q$ by its probability density $q$ with respect some measure $\mu$; so the two notations $S(y;q)$ and $S(y;Q)$ are indistinguishable. 

When working with a parametric model with probability density function $f(y;\theta)$, an important example of proper scoring rules is the log-score, which is defined as $S(y;\theta)=-\log{f(y;\theta)}$ (Good, 1952) and which corresponds to minus the log-likelihood function. 

In this paper we focus on the Tsallis score (Tsallis, 1988), given by
\begin{equation*}
  \label{eq:tsallisscore}
  S(y;\theta) = (\gamma - 1) \int\!  f(y;\theta)^\gamma \, d y - \gamma f(y;\theta)^{\gamma-1},
  \quad \gamma>1 .
\end{equation*}
The Tsallis score gives in general robust procedures (Ghosh and Basu, 2013, Dawid {\em et al.}, 2016), and the parameter $\gamma$ is a trade-off between efficiency and robustness. Applications of the Tsallis score for robust inference has been discussed in, among others, Ghosh and Basu (2013), Pak (2014), Basu {\em et al.} (2016), Ghosh {\em et al.} (2019), and references therein.

%Proper scoring rules can also be extended to the case of a random vector in analogy with composite likelihoods. Let $\{Y_k\}$ be a set of marginal or conditional variables with associated proper scoring rule $S_k$. A proper scoring rule for the random vector $Y$ is defined as
%\begin{equation}
%\label{composite}
%S(\textbf{y};Q)=\sum_{k}S_k(\textbf{y}_k;Q_k),
%\end{equation} 
%where $Y_k \sim Q_k$ when $Y \sim Q$, and $\textbf{y}$ and $\textbf{y}_k$ are the values assumed by $Y$ and $Y_k$, respectively. Scoring rules of the form \eqref{composite} are called composite scoring rules; see Dawid and Musio (2014) and Dawid {\em et al.}\ (2016). Note that when each $S_k$ is the logarithmic score, equation \eqref{composite} is a negative composite log-likelihood; see Varin {\em et al.}\ (2011).  

%%%%%%%%%%%%%%%%%%%%%%%%

\subsection{Inference based on scoring rules} 

The validity of inference about $\theta$ using scoring rules can be justified by invoking the general theory of unbiased $M$-estimating functions. Indeed, inference based on proper scoring rules is a special kind of $M$-estimation (see, e.g., Dawid {\em et al.}, 2016, and references therein). The class of $M$-estimators is broad and includes a variety of well-known estimators. For example  it includes the maximum likelihood estimator (MLE), the maximum composite likelihood estimator (see e.g.\ Varin {\em et al.}, 2011), and robust estimators (see e.g.\ Huber and Ronchetti, 2009 and references therein) among others.

Given a proper scoring rule $S(y;\theta)$, let us denote by $S(\theta)=\sum_{i=1}^n S(y_i;\theta)$ the total empirical score. Moreover,  let $s(y;\theta)$ be the gradient vector of $S(y;\theta)$ with respect to $\theta$, i.e.\ $s(y;\theta)=\partial S(y;\theta)/\partial \theta$. Under broad regularity conditions (see Mameli and Ventura, 2015, and references therein), the scoring rule estimator $\tilde\theta$ is the solution of the unbiased estimating equation 
$$
s(\theta) =  \sum_{i=1}^n s(y_i;\theta) = 0
$$ 
and it is asymptotically normal, with mean $\theta$ and covariance matrix
\begin{eqnarray*}
V (\theta) = K(\theta)^{-1} J(\theta) (K(\theta)^{-1})^{\T}, 
\label{var}
\end{eqnarray*}
where $K(\theta) = E_\theta (\partial s(\theta)/\partial \theta^{\T})$ and $J(\theta) = E_\theta(s(\theta) s(\theta)^{\T} )$ are the sensitivity and the variability matrices, respectively.  The matrix $G(\theta)=V(\theta)^{-1}$ is known as the Godambe information and its form is due to the failure of the information identity since, in general, $K(\theta) \neq J(\theta)$. 

Asymptotic inference on the parameter $\theta$ can be based on the Wald-type statistic
\begin{eqnarray*}
  w_S (\theta) = (\tilde\theta - \theta)^{\T} V(\tilde\theta)^{-1} (\tilde\theta - \theta),
  \label{swe}
\end{eqnarray*}
which has an asymptotic chi-square distribution with $d$ degrees of freedom. In contrast, the asymptotic distribution of the scoring rule ratio statistic
\begin{eqnarray*}
   W_{S}(\theta) = 2 \left\{S(\theta) -S(\tilde\theta)\right\}
  \label{sw}
\end{eqnarray*}
is a linear combination of independent chi-square random variables with coefficients related to the eigenvalues of the matrix $J(\theta)K(\theta)^{-1}$ (Dawid {\em et al.}, 2016). More formally,   
$$
W_{S} (\theta) \, \, \dot\sim \, \, \sum_{j=1}^d \mu_j Z_j^2,
$$
with $\mu_1,\ldots, \mu_d$ eigenvalues of $J(\theta)K(\theta)^{-1}$ and $Z_1, \ldots,Z_d$ independent standard normal variables.  Adjustments of the scoring rule ratio statistic have received consideration in Dawid {\em et al.} (2016). In particular, using the rescaling factor $A(\theta) =(s(\theta)^{\T} J(\theta) s(\theta))/(s(\theta)^{\T} K(\theta) s(\theta))$, we have
\begin{eqnarray*}
W_{S}^{adj} (\theta) = A(\theta) W_{S} (\theta) \, \dot\sim \, \chi^2_d.
\label{swadj}
\end{eqnarray*}

Analogous limiting results can be shown to hold for inference on the scalar parameter $\psi$.  With the
partition $(\psi,\lambda)$, the scoring rule estimating function is similarly
partitioned as $s(y;\theta)=(s_\psi(y;\theta),s_\lambda(y;\theta))$, where
$s_\psi(y;\theta)=(\partial/\partial \psi) S(y;\theta)$ and
$s_\lambda(y;\theta)=(\partial/\partial \lambda) S(y;\theta)$. Moreover,
consider the further partitions
$$
K = \left[
  \begin{array}{cc}
    K_{\psi \psi} & K_{\psi \lambda} \\
    K_{\lambda \psi} & K_{\lambda \lambda}
  \end{array}
\right] \ , \quad K^{-1} = \left[
  \begin{array}{cc}
    K^{\psi \psi} & K^{\psi \lambda} \\
    K^{\lambda \psi} & K^{\lambda \lambda}
  \end{array}
\right] \ ,
$$
and similarly for $G$ and $G^{-1}$. Finally, let $\tilde\lambda_\psi$ be the constrained scoring rule estimate of $\lambda$, let $\tilde\theta_\psi=(\psi,\tilde\lambda_\psi)$, and let $\tilde\psi$ be the $\psi$ component of $\tilde\theta$. 

A profile scoring rule Wald-type statistic for the $\psi$ component
may be defined as
$$
 w_{Sp}  (\psi) = (\tilde\psi - \psi) (\tilde{G}^{\psi \psi})^{-1/2},
$$
and it has an asymptotic $N(0,1)$ null distribution. Similarly, the profile scoring rule score-type statistic
$s_\psi (\tilde\theta_{\psi})^T K^{\psi \psi}
(G^{\psi \psi})^{-1} K^{\psi \psi} s_\psi (\tilde\theta_{\psi})$ has an asymptotic $\chi^2_1$ null distribution. Finally, we have that the asymptotic distribution of the profile scoring rule ratio
statistic for $\psi$, given by $W_{Sp} (\psi) = 2 \left( S(\tilde\theta_{\psi}) - S(\tilde\theta) \right)$,
is $ \nu \chi^2_1$, where $\nu = (\tilde{K}^{\psi \psi})^{-1} \tilde{G}^{\psi \psi}$.  In view of this, an adjusted profile scoring rule ratio statistic can be computed as
  \begin{eqnarray*}
   W_{Sp}^{adj} (\psi) = \frac{W_{Sp} (\psi)}{\nu}\, \, \dot\sim \,\, \chi^2_1.
    \label{swadj}
  \end{eqnarray*}
The adjusted profile scoring rule root, analogous to (\ref{r}), can be defined as
\begin{eqnarray*}
r_{Sp} (\psi) = \text{sign} (\tilde\psi - \psi) \sqrt{ W_{Sp}^{adj} (\psi) },
\label{rsradj}
\end{eqnarray*} 
which has an asymptotic standard normal distribution.

\subsection{Examples}

In this section, the Tsallis scoring rule is illustrated for two well-known models: the regression model and the exponential family. All the quantities necessary to compute $w_{Sp}  (\psi)$ and $r_{Sp} (\psi)$ are derived.

\vspace{0.5cm}
 
\noindent {\bf Tsallis score for regression.} Consider the general regression model
\begin{eqnarray}
y_i= \mu(x_i,\beta)+\varepsilon_i, \quad \text{with} \, \, i=1,\ldots, n,
\label{modello}
\end{eqnarray}
with $x_i$ vector of fixed covariates, $\beta$ an unknown $p$-dimensional parameter, and $\varepsilon_i$ independent and identically distributed $N(0,\sigma^2)$ random variables. The classical linear model is obtained with $\mu(x_i,\beta)= x_i^{\T} \beta$, $i=1,\ldots,n$.

For this model, the total Tsallis score for $\theta=(\beta,\sigma^2)$ is 
\begin{eqnarray}
S(\theta) = \sum_{i=1}^n\left[-\gamma\left(\frac{1}{\sqrt{2\pi\sigma^2}}\right)^{(\gamma-1)}\exp{\left(-\frac{(\gamma-1)}{2\sigma^2}(y_i-\mu_i(x_i,\beta))^2\right)} +c_\gamma\right],
\label{tnl}
\end{eqnarray}
with $c_\gamma= (\gamma-1)/\sqrt{\gamma}(2\pi\sigma^2)^{\frac{(\gamma-1)}{2}}$.
It is possibile to show that (Girardi {\em et al.}, 2020)
\begin{equation*}
K(\theta)= 
\begin{pmatrix}
\frac{\xi_{\alpha}}{n}\frac{\partial \mu}{\partial\beta}^T\frac{\partial \mu}{\partial\beta}& 0 \\
0 & \varsigma_{\alpha}\\
\end{pmatrix},
\end{equation*}
where $\frac{\partial \mu}{\partial\beta}^T=(\frac{\partial \mu_1}{\partial\beta},\cdots,\frac{\partial \mu_n}{\partial\beta})$ is a $p\times n$ matrix, with $\mu_i=\mu(x_i,\beta)$, $i=1,\ldots,n$, and $\xi_{\alpha}$ and $\varsigma_{\alpha}$ are the same as given in 
 Gosh and Basu (2013) for the linear regression model (see Sect.6), namely 
 $\xi_{\alpha}=(2\pi)^{-\alpha/2}\sigma^{-(\alpha+2)/2}(1+\alpha)^{-3/2}$ and $\varsigma_{\alpha}=\frac{1}{4}(2\pi)^{-\alpha/2}\sigma^{-(\alpha+4)/2}\frac{2+\alpha^2}{(1+\alpha)^{5/2}}$. Moreover 
\begin{equation*}
J(\theta) =
\begin{pmatrix}
\frac{\xi_{2\alpha}}{n}\frac{\partial \mu}{\partial\beta}^T\frac{\partial \mu}{\partial\beta}& 0 \\
0 & \varsigma_{2\alpha}-\frac{\alpha^2}{4}\xi_{\alpha}\\
\end{pmatrix}.
\end{equation*}
For the Tsallis score (\ref{tnl}), it can be shown that the IF is bounded in $y$ for all $\gamma>1$ (Girardi {\em et al.}, 2020).

%%%%%%%%%%%%%%%%%%%%%%%%%%%%%%%%%%%%%

\vspace{0.5cm}

\noindent {\bf  Tsallis score for exponential family.} Let $Y$ have a distribution belonging to the canonical exponential family
\begin{equation}
\label{expo}
f(y; \theta)=e^{\sum_{i=1}^s\theta_i t_i(y)-c(\theta)+d(y)},
\end{equation}
where $c(\theta)$ is a strictly convex $C^{\infty}$ function, $\theta$ is the $s$-dimensional natural parameter, and $t(y)=(t_1(y),\ldots, t_s(y))$ is a vector of  sufficient statistics. It can be shown that 
$$
\int\!  f(y;\theta)^\gamma \, d y =e^{c(\gamma \theta)-\gamma c(\theta)}E_{\theta}\Big[e^{(\gamma-1)d(Y)}\Big]
$$ (see Nielsen and Nock, 2012).
 
The Tsallis score can be written as
\begin{equation}
\label{expo-Tsallis}
S(y;\theta)=(\gamma-1)e^{c(\gamma \theta)-\gamma c(\theta)}E_{\theta}\Big[e^{(\gamma-1)d(Y)}\Big]-\gamma f(y;\theta)^{(\gamma-1)}.
\end{equation}
In particular, when $d(y)$ is equal to $0$ (as in the Normal, Gamma and Beta models 
 for instance), (\ref{expo-Tsallis}) becomes a closed-form formula since 
$E_{\theta}\Big[e^{(\gamma-1)d(y)}\Big]=E_{\theta}[1]=1$ and reduces to 

\begin{equation}
\label{expo1-Tsallis}
S(y;\theta)=(\gamma-1)e^{c(\gamma \theta)-\gamma c(\theta)}-\gamma f(y;\theta)^{(\gamma-1)}.
\end{equation}
For instance, if $Y \sim \Gamma(\alpha, \beta)$, we have
$$f(y;\alpha,\beta)=e^{ -y \beta +(\alpha-1)\ln y+ \alpha \ln \beta -\ln \Gamma(\alpha)}, $$
with $\theta=(\alpha-1, -\beta)$ and $$c(\theta)=\ln \Gamma(\alpha)-\alpha \ln \beta=\ln \Gamma(\theta_1+1)-(\theta_1+1) \ln |-\theta_2|.$$
Substituting (\ref{expo}) in (\ref{expo1-Tsallis}), we find the following expression for the Tsallis score
\begin{equation}
\label{normal-Tsallis}
S(y;\theta)=(\gamma-1)e^{c(\gamma \theta)-\gamma c(\theta)}-\gamma e^{(\gamma-1)(\theta t(y)^T-c(\theta))}.
\end{equation}
Define $c_i(\theta) = \partial c(\theta)/ \partial(\theta_i)$ (note that $c_i(\theta) = E_{\theta}( t_i(y) )$), we have 
\begin{eqnarray}
s(y;\theta_i)&=&\frac{\partial S(y;\theta)}{\partial \theta_i }=\gamma(\gamma-1)\Big[e^{c(\gamma \theta)-\gamma c(\theta)}( c_i(\gamma \theta) - c_i(\theta) )) + \\
&-& e^{(\gamma-1)(\sum_{i=1}^s\theta_i t_i(y)-c(\theta)) } \Big(t_i(y)-c_i(\theta) \Big) \Big].
\end{eqnarray}
The Tsallis  estimator $\tilde{\theta}$ is the solution of the system
$$
\sum_{j=1}^n s(y_j; \theta_i)=0 \quad i=1, \cdots,s.
$$
The estimator is robust if and only if  $s(y;\theta_i), i=1, \ldots,s,$ is bounded in $y$  for all $\theta$'s. In our case this condition requires that  
$$
 e^{(\gamma-1)(\sum_{i=1}^s\theta_i t_i(y)-c(\theta)) } \Big(t_i(y)- E_{\theta}( t_i(y) ) \Big) 
$$  
is a bounded function of $y$ for each $i=1, \cdots,s$ and for each value of  $\theta=(\theta_1,\ldots,\theta_s)$.
If $\gamma >1$, these conditions are satisfied in the normal  model, in the Gamma model if $\alpha >1$ and  in the Beta model if $\alpha$ and $\beta$ are both $>1$.

%In particular, using the asymptotic expansions of the mean and the variance of $r_{sr} (\psi)$, it is possible to defined the modified profile signed scoring rule root statistic 
%\begin{eqnarray}
%r_{sr}^M (\psi) = \frac{r_{sr} (\psi) - m_p(\psi)}{\sqrt{\nu + v_p(\psi)}},
%\label{rsradj2}
%\end{eqnarray}
%where $m_p(\psi)$ is of order $O(n^{-1/2})$ and $v_p(\psi)$ is of order $O(n^{-1})$.  If third- and higher-order cumulants of $r_{sr}^M (\psi)$ are of order $O(n^{-3/2})$ or smaller, then the distribution of the modified profile signed scoring rule root statistic $r_{sr}^M (\psi)$ follows asymptotically a normal distribution with error $O(n^{-3/2})$. This happens in models with all higher cumulants zero and for the logarithmic score. In general, if such conditions are not fulfilled, $r_{sr}^M (\psi)$ is asymptotically normal distributed with error $O(n^{-1})$. The analytical expressions of the mean and variance corrections are derived in Mameli and Ventura (2015) and they involve several expected values of scoring rules derivatives,  whose computation may be cumbersome even in simple models. Alternatively, a parametric bootstrap approach, that avoids onerous calculations specific of analytical adjustments, can be used as illustrated in Mameli {\em et al.} (2017).

%%%%%%%%%%%%%%%%%%%%%%%%%%%%%%%%%%%%%%%%%
%%%%%%%%%%%%%%%%%%%%%%%%%%%%%%%%%%%%%%%%%
%%%%%%%%%%%%%%%%%%%%%%%%%%%%%%%%%%%%%%%%%
%%%%%%%%%%%%%%%%%%%%%%%%%%%%%%%%%%%%%%%%%

\section{Confidence distributions from scoring rules} 

In this section we discuss how to derive CDs from proper scoring rules. In particular, we discuss asymptotic CDs based on first-order approximations of SR pivotal quantities. 

Paralleling results in Section 2.1 for likelihood based CDs, a recipe to derive an asymptotic CD from scoring rules is based on pivotal quantites. Let $q_S (\psi;y)$ a scoring rule pivotal quantity, such as the profile Wald-type statistic $w_{Sp}  (\psi)$ or the adjusted profile scoring rule root $r_{Sp} (\psi)$.

Thus, 
\begin{eqnarray}
C_{S}^w(\psi) \, \dot{=} \, \Phi\left( (\psi - \tilde\psi) (\tilde{G}^{\psi \psi})^{-1/2} \right)
\label{cd1}
\end{eqnarray}
and
\begin{eqnarray}
C_{S}^r(\psi) \, \dot{=} \, \Phi \left( \text{sign} (\psi - \tilde\psi) \sqrt{ W_{Sp}^{adj} (\psi) } \right)
\label{cd2}
\end{eqnarray}
are first-order asymptotic CDs. The corresponding CCs are, respectively,
$$
cc_{S}^w(\psi) \, \dot{=} \, \frac{\phi \left((\psi - \tilde\psi) (\tilde{G}^{\psi \psi})^{-1/2} \right)}{\sqrt{G^{\psi \psi}}}
$$
and
$$
cc_{S}^r(\psi)  \, \dot{=} \, \phi \left( \text{sign} (\psi - \tilde\psi) \sqrt{ W_{Sp}^{adj} (\psi) } \right) \,  \left| \frac{\partial W_{Sp}^{adj} (\psi)^{1/2}}{\partial \psi} \right|,
$$
where $\phi(\cdot)$ is the density function of the standard normal distribution.
As for likelihood based CDs,  (\ref{cd2}) can reflect asymmetry, unlike the simpler Wald-type confidence distribution (\ref{cd1}). 

For instance, using (\ref{cd2}), the confidence median is $\tilde\psi$ and an $(1-\alpha)$ equi-tailed confidence intervals can be obtained as $\{\psi : |r_{Sp} (\psi)| \leq z_{1-\alpha/2} \}$. When testing, for instance, $H_0: \psi = \psi_0$ against $H_1: \psi < \psi_0$, the $p$-value is $p=C_{S}^r(\psi_0)$, while when testing
 $H_0: \psi = \psi_0$ against $H_1: \psi \neq \psi_0$ the $p$-value is $p = 2(1-\Phi(|r_{Sp}(\psi_0)|))$. A measure of evidence for $\psi_1<\psi<\psi_2$ can be computed as $C_{S}^r(\psi_2)-C_{S}^r(\psi_1)$.

Finally, note  that (\ref{cd1}) coincides with the asymptotic first-order Bayesian posterior for $\psi$ discussed in Giummol\'e {\em et al.} (2019) and in Pauli {\em et al.} (2011), and that, when in particular $S(\theta)$ is the logarithmic score, (\ref{cd1}) reduces to (\ref{cdwald}).

%%%%%%%%%%%%%%%%%%%%%%%%%%%
%%%%%%%%%%%%%%%%%%%%%%%%%%%

\subsection{Robustness of the tail area}

From the general theory of $M$-estimators, the influence function ($IF$) of the estimator $\tilde\theta$ is given by
\begin{eqnarray}
IF (y;\tilde\theta) = K(\theta)^{-1} s(y;\theta),
\label{ifsco}
\end{eqnarray}
and it measures the effect on the estimator $\tilde\theta$ of an infinitesimal contamination at the point $y$, standardised by the mass of the contamination. The estimator $\tilde\theta$ is B-robust if and only if $s(y;\theta)$ is bounded in $y$. Note that the $IF$ of the MLE is proportional to the score function; therefore, in general, MLE has unbounded $IF$, i.e. it is not B-robust. The general theory of robust tests has been discussed in Heritier and Ronchetti (1994). \\
Sufficient conditions for the robustness of the Tsallis score are discussed, for instance, in Basu {\em et al.} (1998) and Dawid {\em et al.} (2016). 
In this section we investigate the effects of model deviations on CDs. In this respect, let us write the scoring rule pivotal quantity more generally as $q_S (\psi; T(\hat{F}_n))$, where $\hat{F}_n$ is the empirical distribution function and $T(F)$ is the functional defined by the scoring rule estimating equation $\int s(y;T(F)) \, dF(y)=0$, where $F=F(y;\theta)$ is the assumed parametric model. Indeed, both the scoring rule pivotal quantities $w_{Sp}(\psi)$ and $r_{Sp}(\psi)$ are functions of the scoring rule estimator.  
For CD  inference  it is of interest the tail area. 
For a fixed value of $\psi$ the CD tail area is given by
%For a fixed value of $\psi$, let us focus on the CD tail area 
\begin{eqnarray*}
C_S(\psi)  = \Phi (q_S (\psi; T(\hat{F}_n)).
\label{tailarea}
\end{eqnarray*}

To study the effects of deviations from the assumed parametric model $F$, we use the tail area influece function (see, e.g., Field and Ronchetti,1990, and Ronchetti and Ventura, 2001)
\begin{eqnarray}
TAIF(y;T) = \left. \frac{\partial}{\partial \varepsilon} \Phi (q_S (\psi; T(F_\varepsilon))) \right|_{\varepsilon=0},
\label{taif}
\end{eqnarray}
where $F_\varepsilon = (1-\varepsilon) F + \epsilon \Delta_y$ and $\Delta_y$ is the probability measure which puts mass 1 at the point $y$. The $TAIF(y;T)$ thus describes the normalized influence on the CD tail area of an infinitesimal observation at $y$ and,  by considering its supremum, it can be used to evaluate the maximum bias of the tail area on the $\varepsilon$-neighborhood of $F$. 

By computing (\ref{taif}) we can identify the functional,
defined by $s(y;T(F))$, for which (\ref{tailarea}) is robust in the sense that the tail area influence function is bounded and therefore the maximum bias of the corresponding $p$-value is bounded in the $\varepsilon$-neighborhood of the model. After some calculations, we obtain
\begin{eqnarray}
TAIF(y;T) & = &  \phi(q_S(\psi;T(F))) \, \, \frac{\partial q_S(\psi;T(F))}{\partial T(F)} \, \, \left. \frac{\partial T(F_\varepsilon)}{\partial \varepsilon} \right|_{\varepsilon=0},
\label{taif2}
\end{eqnarray}
where the last term in (\ref{taif2}) is the IF (\ref{ifsco}) of the scoring rule estimator. Thus, the tail area influence function for the CD tail area at the statistical model $F$ is proportional to the scoring rule estimating function and this gives an immediate  handle on robustness. Furthermore, it  is bounded with respect to $y$ when the scoring rule estimating function is bounded.

When considering robust scoring rules, i.e.\ scoring rules that lead to estimators with bounded IF, such as the Tsallis scoring rule (see, e.g., Dawid {\em et al.}, 2016), then both (\ref{cd1}) and (\ref{cd2}) are robust CDs. On the contrary, when considering the  logarithmic score, typically (\ref{taif2}) is not bounded since the MLE is not B-robust.

%%%%%%%%%%%%%%%%%%%%%%%%%%%%%%%%%%%

%%%%%%%%%%%%%%%%%%%%%%%%%%%%%%%%%%%%%%%%%
%%%%%%%%%%%%%%%%%%%%%%%%%%%%%%%%%%%%%%%%%
%%%%%%%%%%%%%%%%%%%%%%%%%%%%%%%%%%%%%%%%%
%%%%%%%%%%%%%%%%%%%%%%%%%%%%%%%%%%%%%%%%%

\section{Classical case studies in clinical research and simulation studies} 

In this section the practical usage of robust CDs will be presented in some classical contexts, both by simulated and real-life data.

In particular, we consider the following frameworks: a  regression analysis, the comparison of two means in the presence of heteroschedasticity and inference on the AUC.

%%%%%%%%%%%%%%%%%%%%%%%%%%%%%%%%%%%%%%%%%
%%%%%%%%%%%%%%%%%%%%%%%%%%%%%%%%%%%%%%%%%

\subsection{Two sample comparison}

Many experimental measurements are reported as realizations from a normal distribution, and the simplest comparison we can make is between two groups. The independent samples $t$-test is used when two separate sets of independent and identically distributed samples are obtained, one from each of the two populations being compared.

Let us assume that data are heteroscedastic, i.e.\ that standard deviations are different from each other. In particular, let us assume that $x=(x_1,\ldots,x_{n_x})$ and that $y=(y_1,\ldots,y_{n_y})$ are two random samples from, respectively, $X \sim N(\mu_x,\sigma^2_x)$ and $Y \sim N(\mu_y,\sigma^2_y)$.  Writing $\theta=(\mu_x,\mu_y,\sigma^2_x,\sigma^2_y)$, the Tsallis empirical score is
$$
S(\theta) = (\gamma-1) c_x + \gamma \sum_{i=1}^{n_x}  f_x(x_i;\mu_x,\sigma^2_x)^{\gamma-1} +
(\gamma-1) c_y + \gamma \sum_{i=1}^{n_y}  f_y(y_i;\mu_y,\sigma^2_y)^{\gamma-1},
$$
where $c_x$ and $c_y$ are suitable constants, depending on $\sigma^2x$ and $\sigma^2_y$, respectively. Typically, the parameter of interest is $\psi=\mu_x-\mu_y$. 

\vspace{0.2cm}

\noindent {\bf Simulation results.} In order to assess the quality of CD inference for $\psi$ based on the Tsallis scoring rule in comparison to likelihood based CD inference, we ran a simulation experiment in which the robustness constant $\gamma$ is fixed in such a way that the resulting estimator is 10\% less efficient than the MLE, under the true model. We generated $10^5$ datasets with sizes of the two samples $(n_1=10,n_2=20)$ form the true model with parameter values $\mu_x=2,\mu_y=0,\sigma_x=\sigma_x=1$, i.e. $\psi=2$, with and without contamination. Contaminated data were generated shiftwise, by adding -7 to the last observation of the first sample, i.e. the sample with size $n_1$. For each dataset we computed the coverage of Wald- and $r_p$-type CD intervals at various confidence levels. Furthermore, the uniformity of the $p$-values when testing $H_0:\psi=2$ against $H_1:\psi<2$ is also checked.

\begin{figure}
\includegraphics[width=0.44\textwidth]{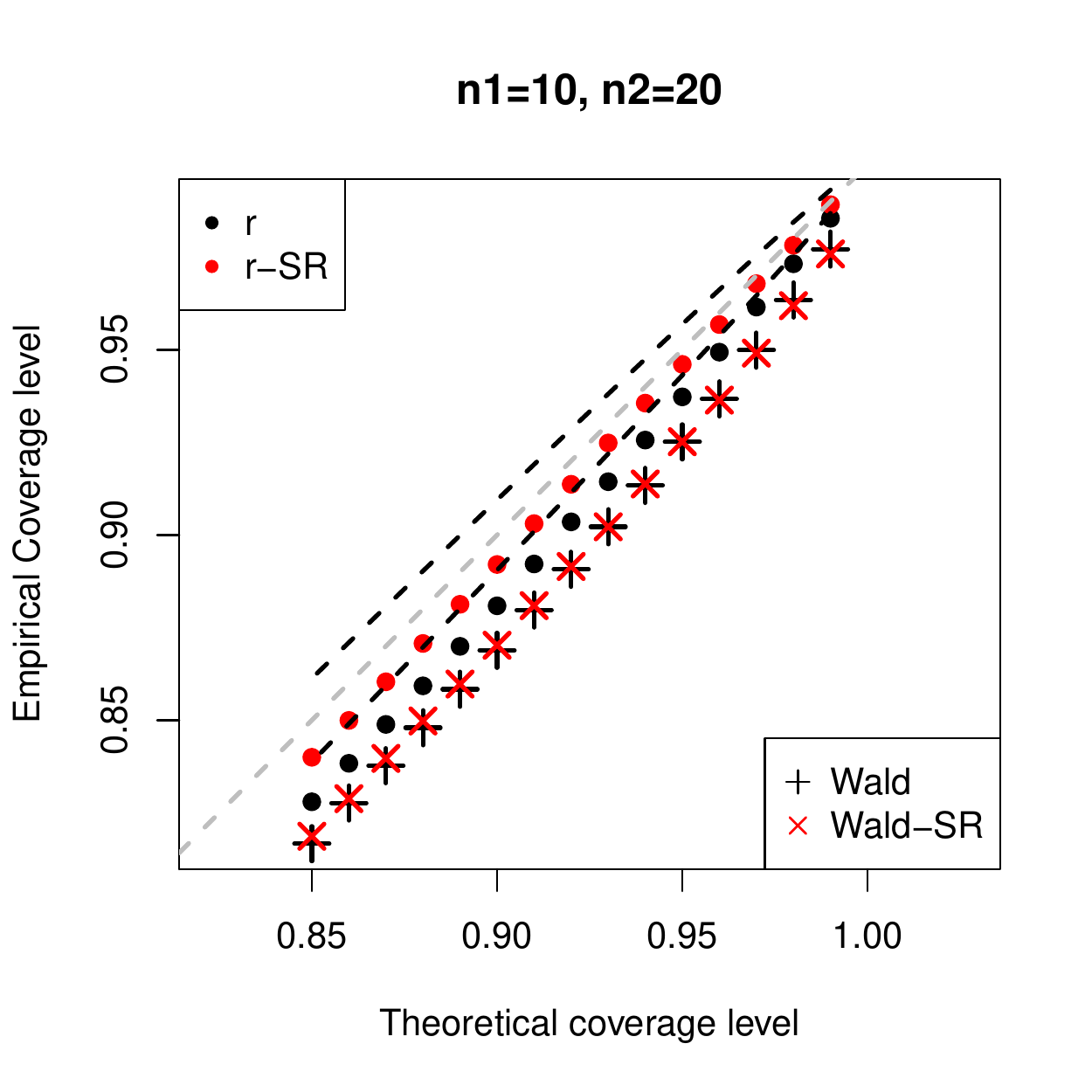}
\includegraphics[width=0.44\textwidth]{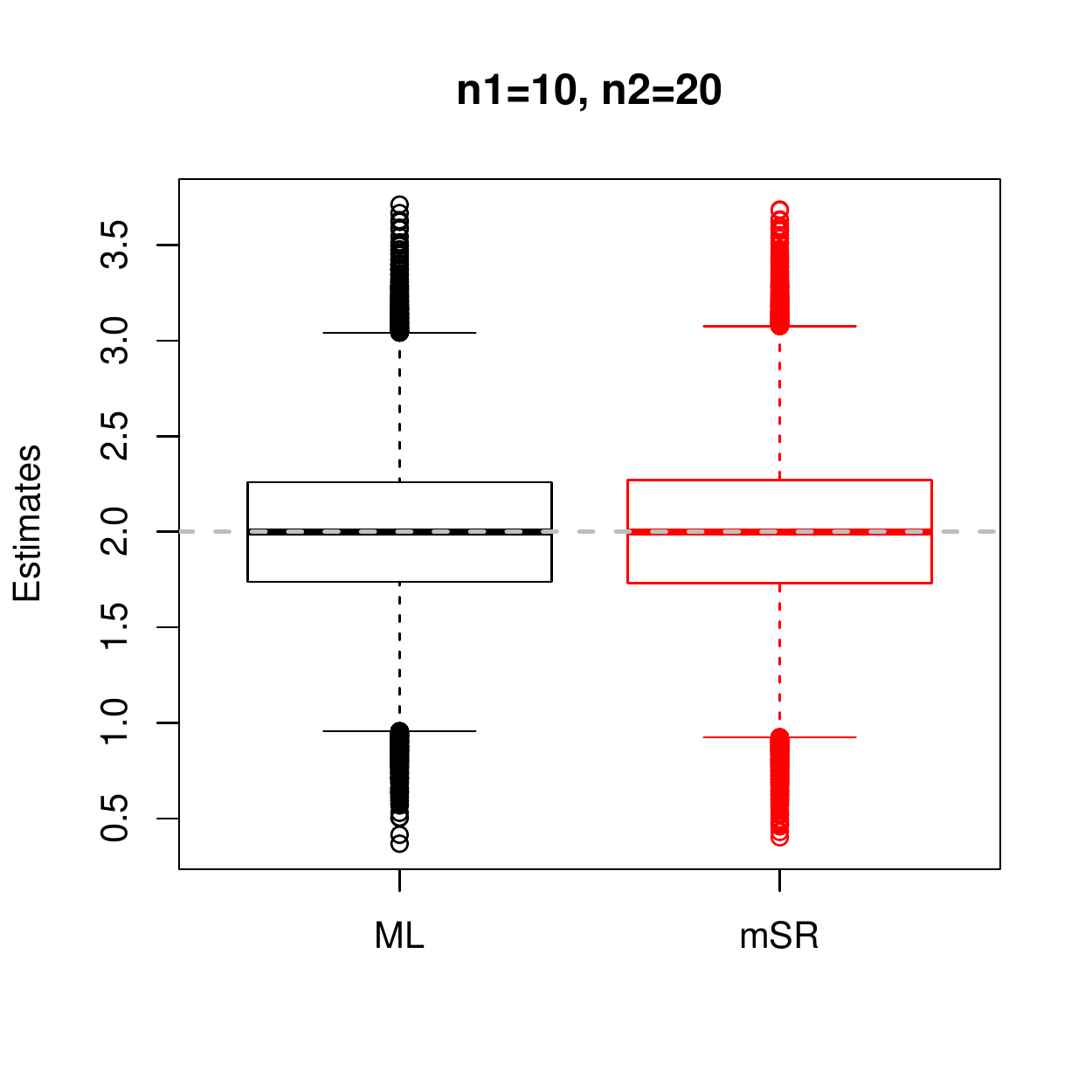}\\
\includegraphics[width=0.44\textwidth]{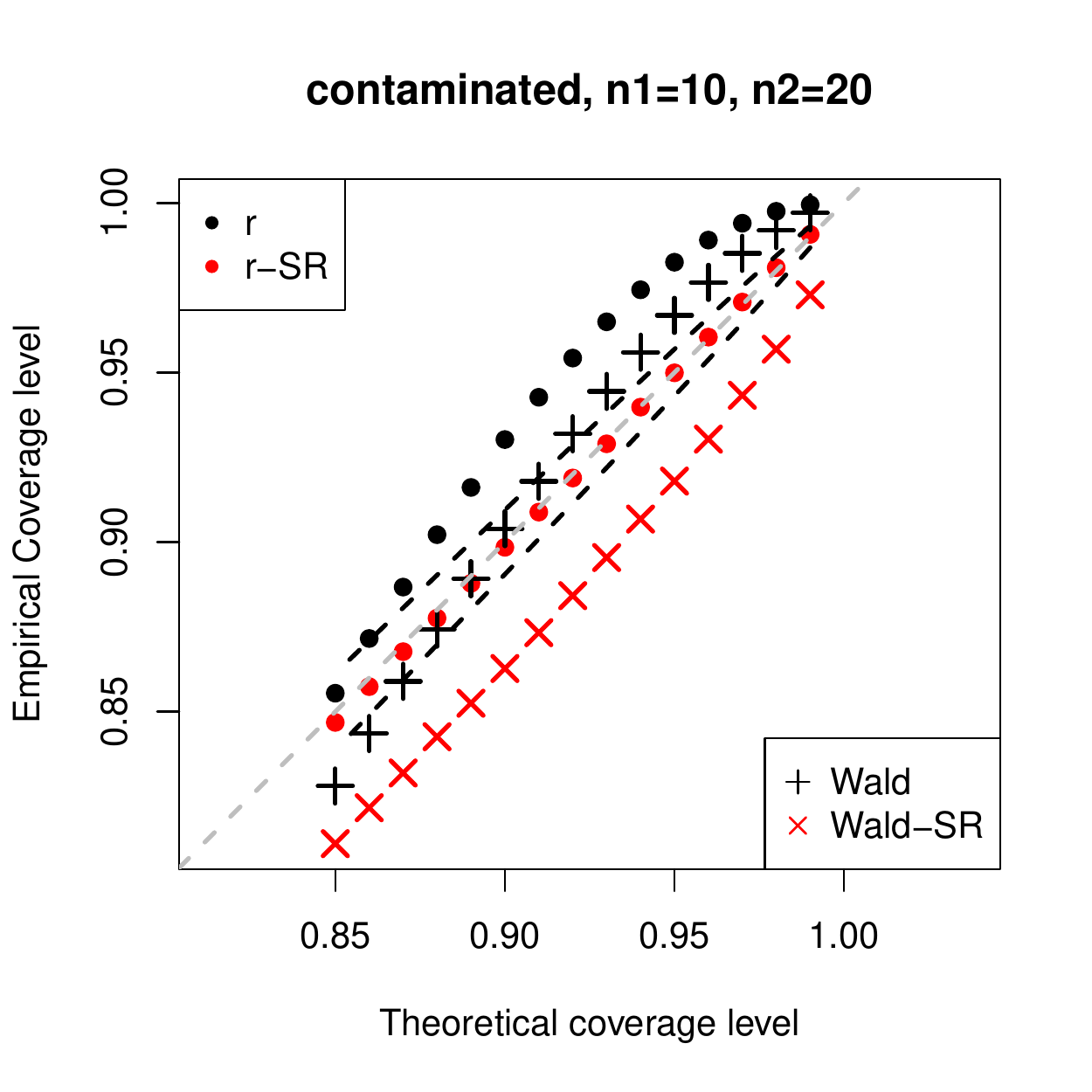}
\includegraphics[width=0.44\textwidth]{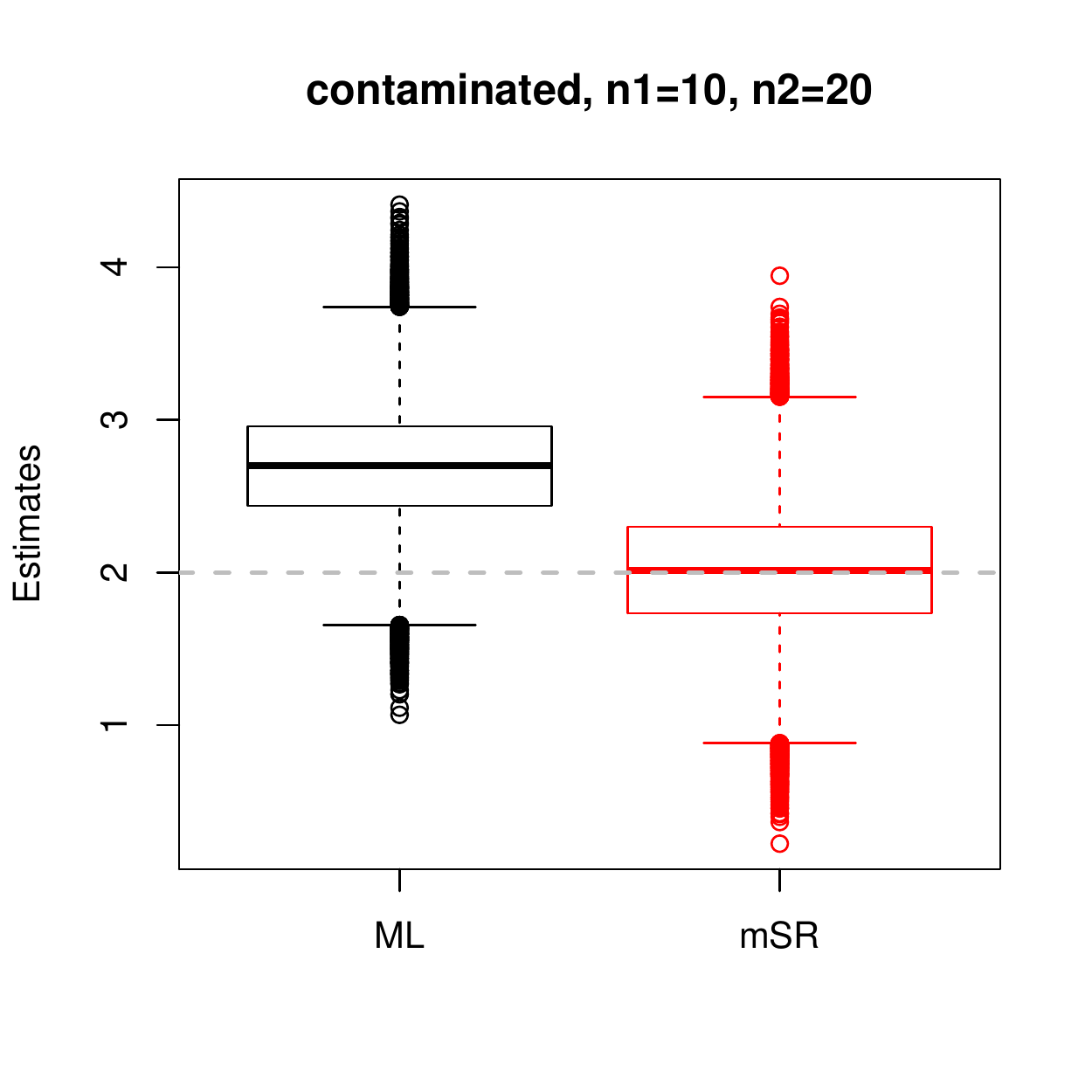}\\
\caption{\small Two normal random variables. Empirical coverage of CIs and distribution of the CD estimates of $\psi$ with data generated under the true model without contamination (first row) and with contamination (second row). Dashed lines represent 10 $\times$ Monte Carlo standard error from the theoretical confidence level.}\label{fig:ttest-sim-1}
\end{figure}

\begin{figure}
\includegraphics[width=0.24\textwidth]{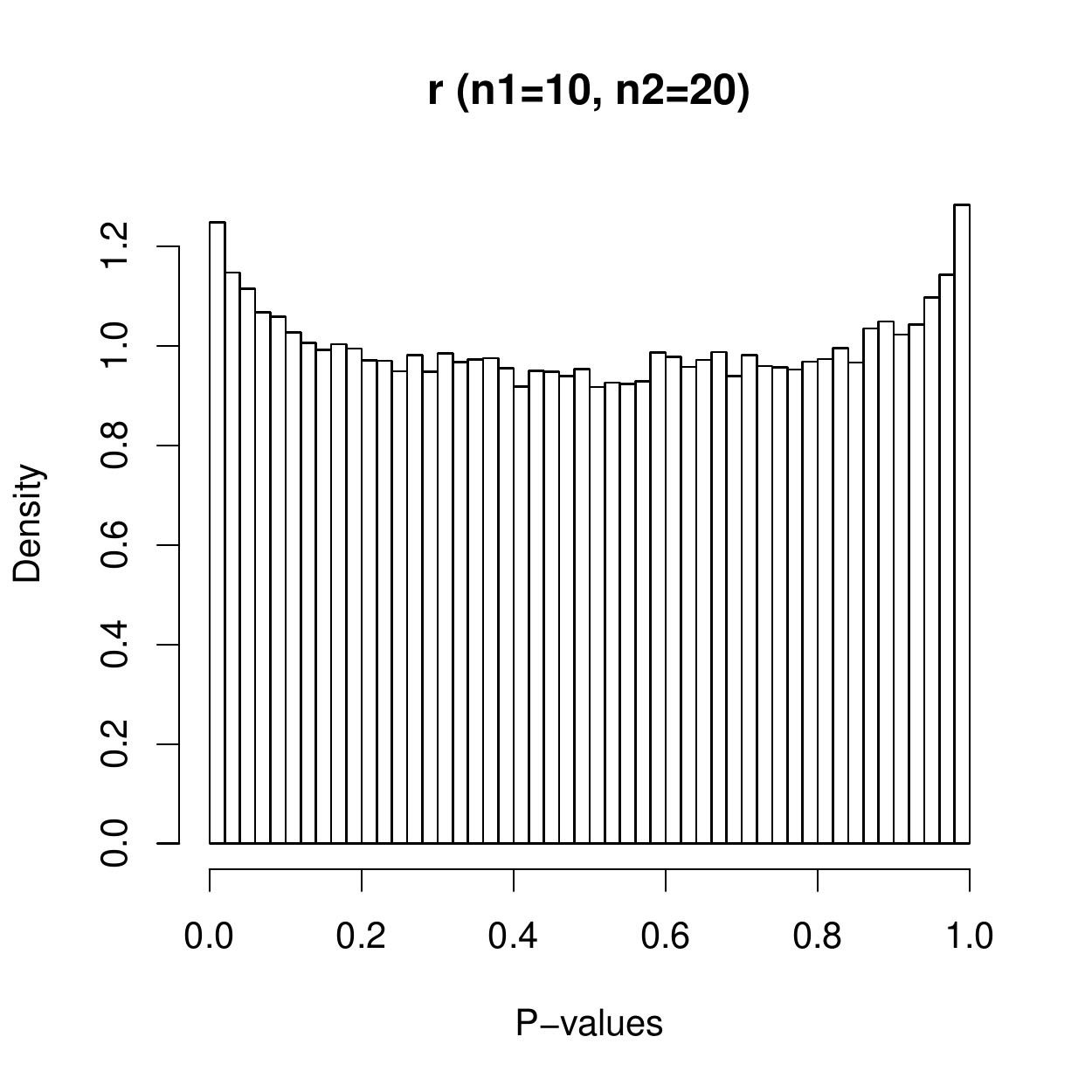}
\includegraphics[width=0.24\textwidth]{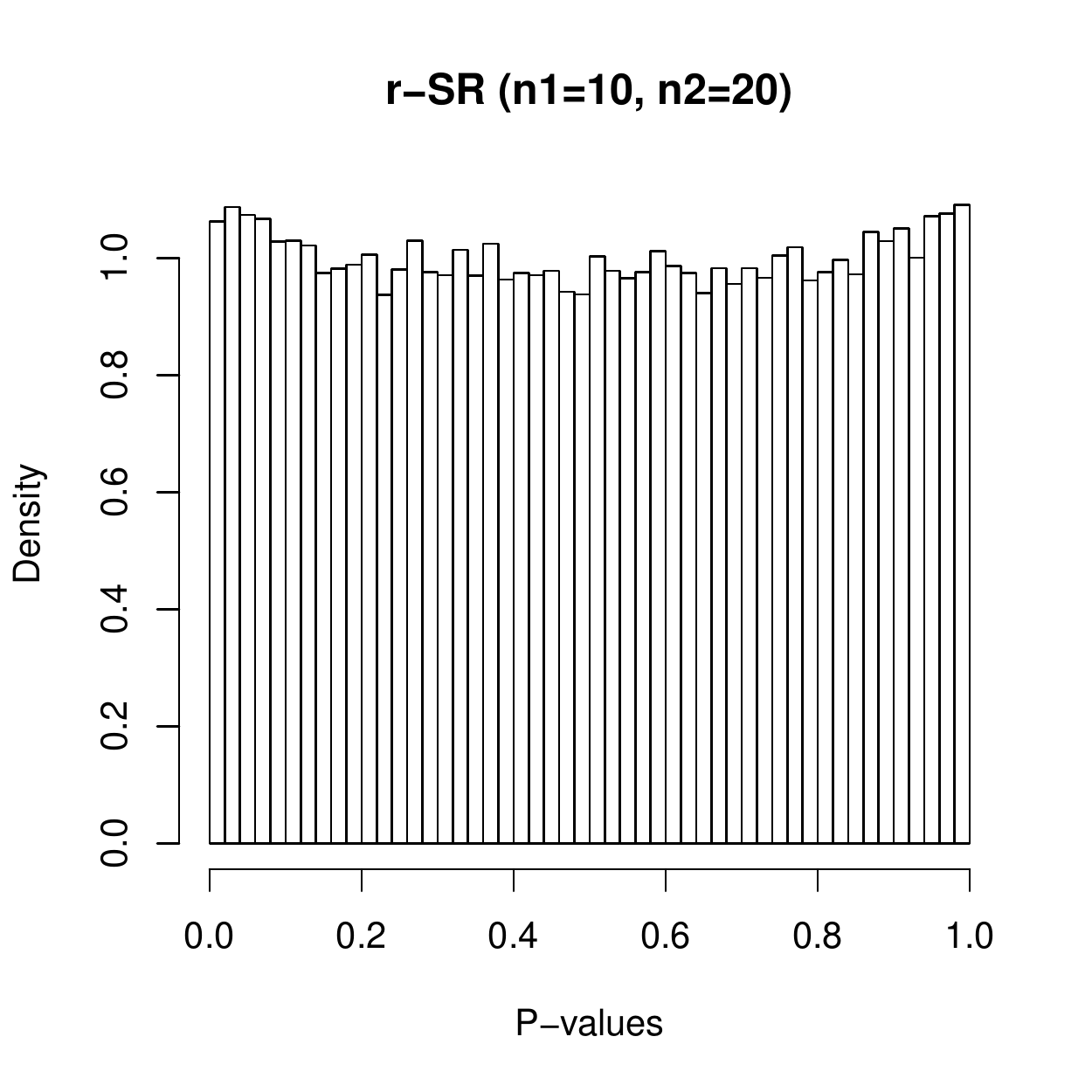}
\includegraphics[width=0.24\textwidth]{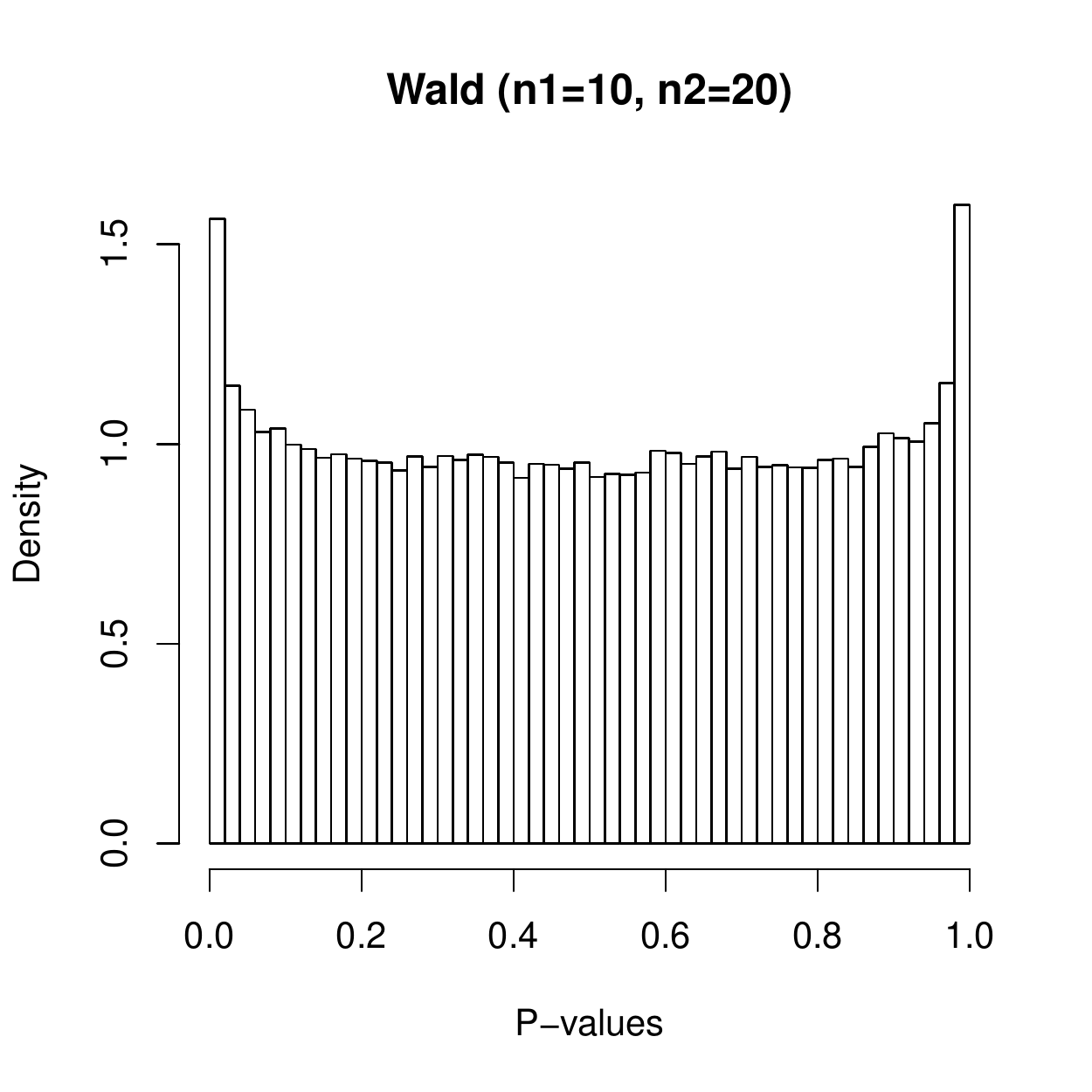}
\includegraphics[width=0.24\textwidth]{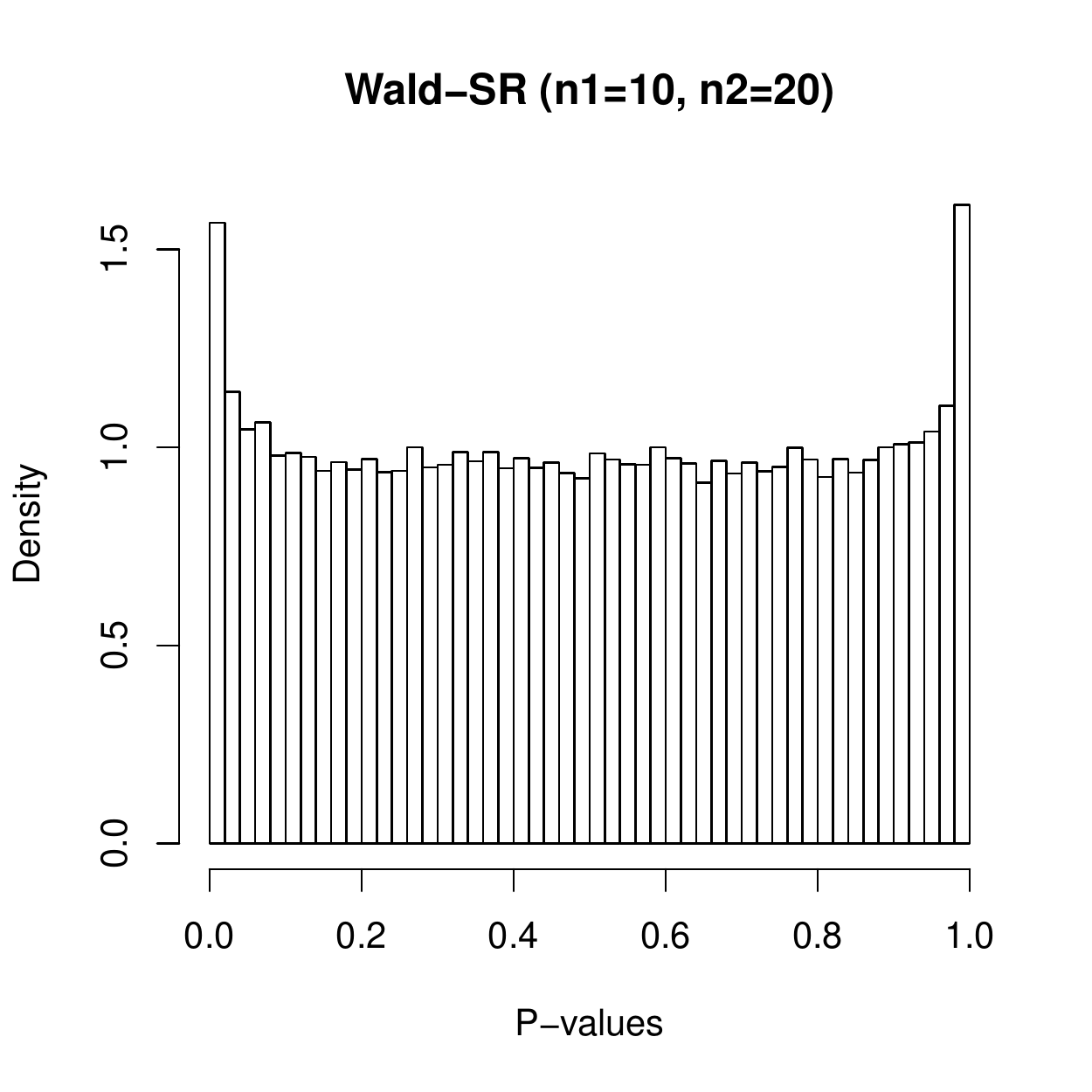}
\includegraphics[width=0.24\textwidth]{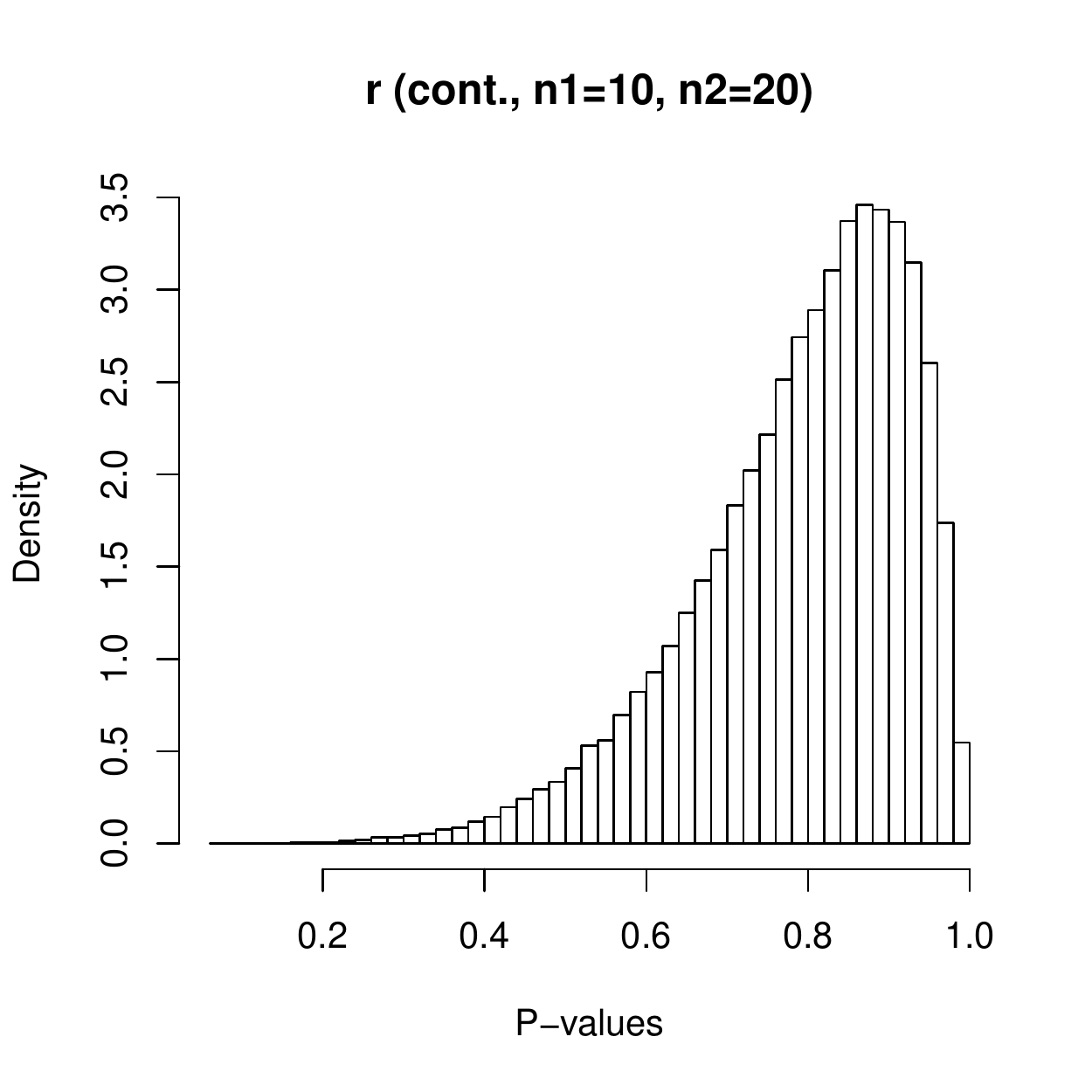}
\includegraphics[width=0.24\textwidth]{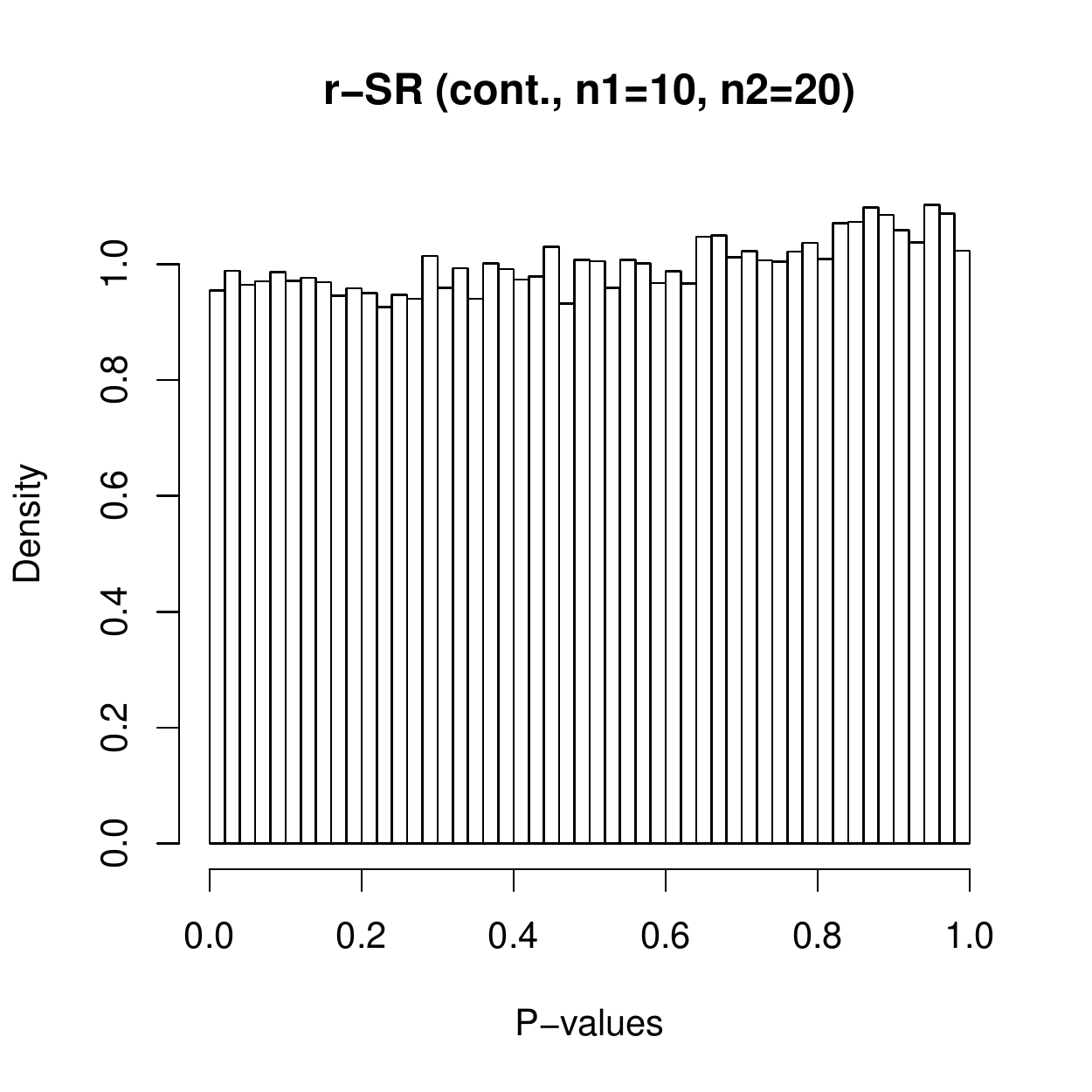}
\includegraphics[width=0.24\textwidth]{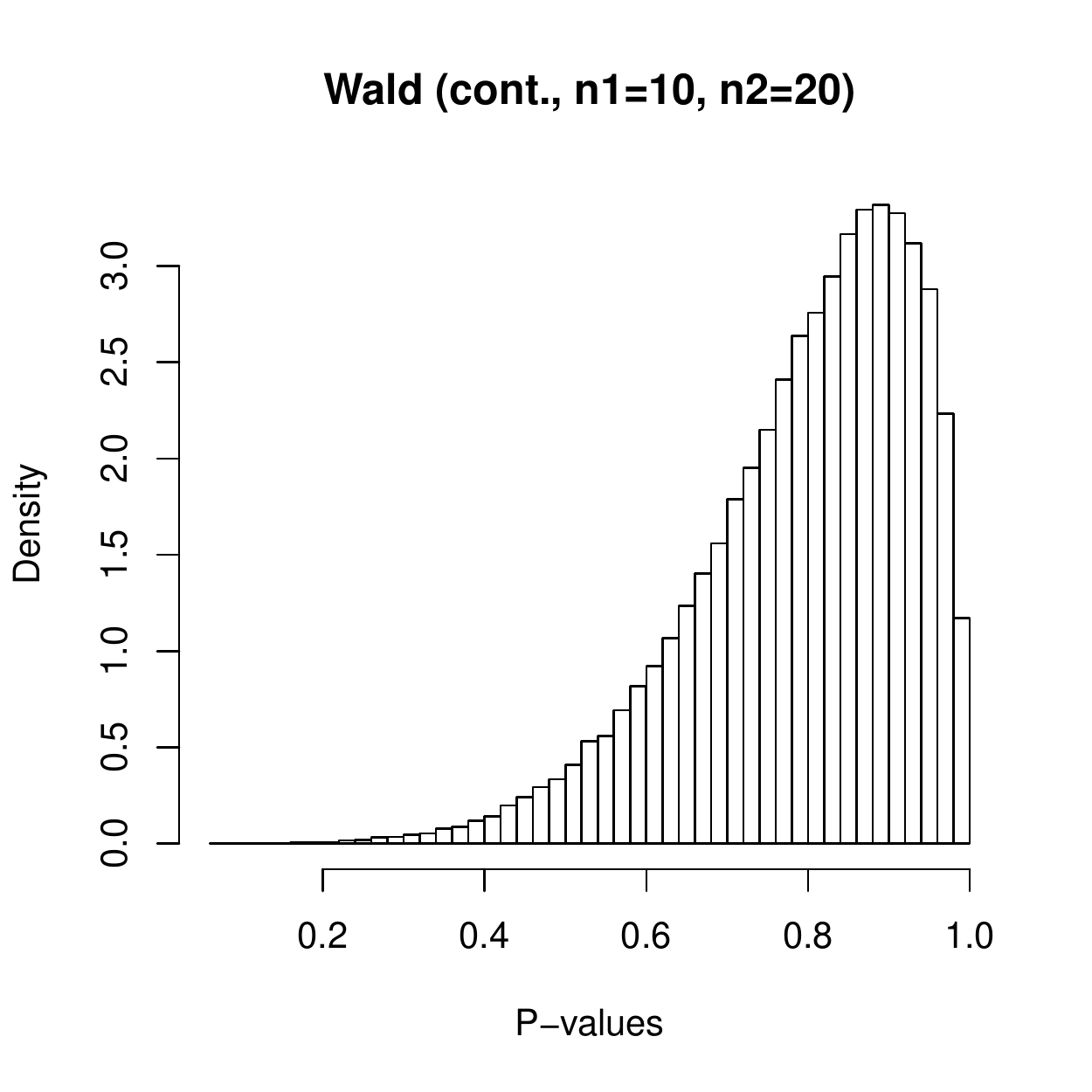}
\includegraphics[width=0.24\textwidth]{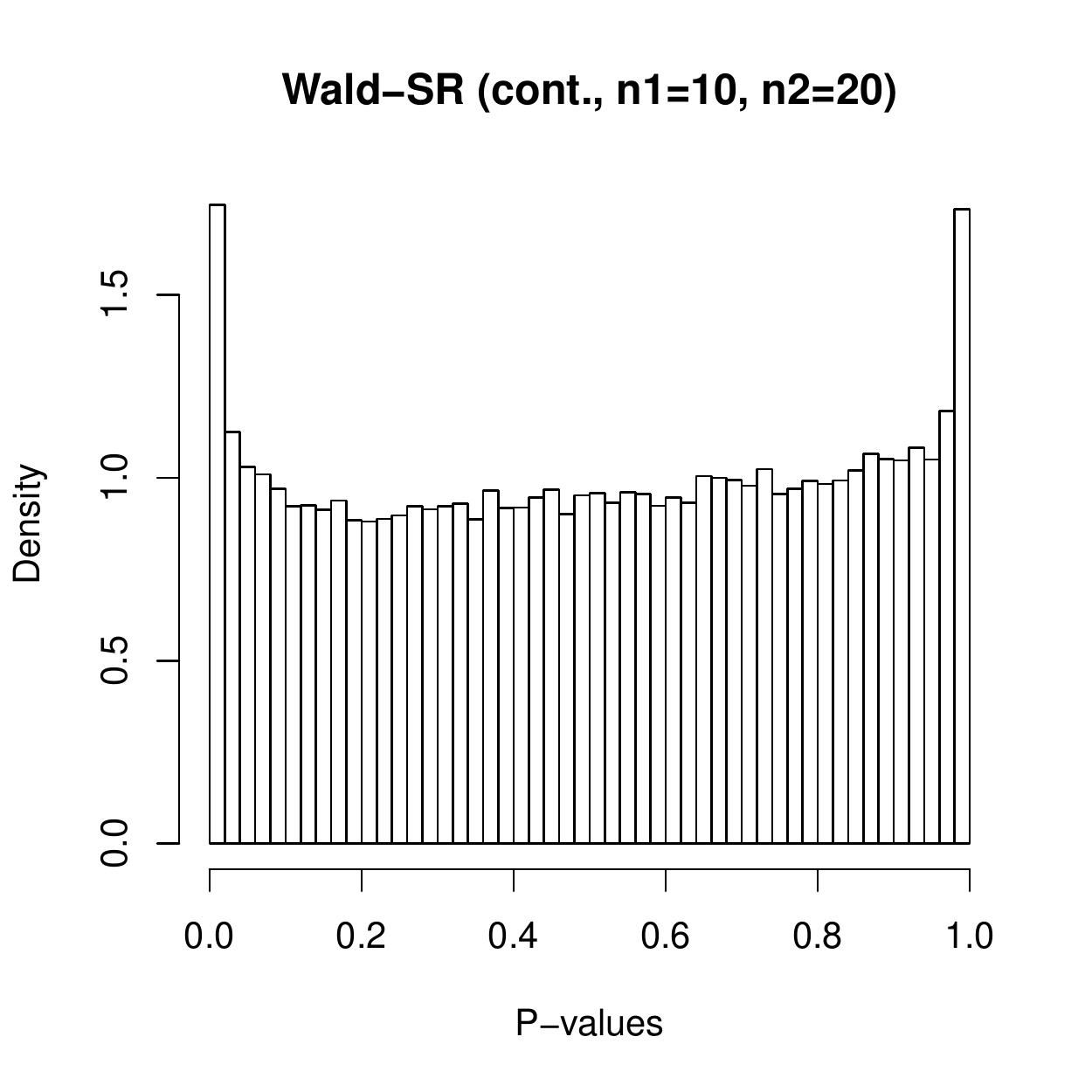}
\caption{\small Two normal random variables. Empirical distribution of $p$-values for $H_0:\psi = 2$ against $H_1:\psi < 2$ using either non contaminated data (first row) or contaminated (second row).}\label{fig:ttest-sim-2}
\end{figure}

From the simulation results shown in Figure~\ref{fig:ttest-sim-1} and \ref{fig:ttest-sim-2}  % reports the empirical coverages of bilateral confidence intervals based on four CDs from the pivotal quantities: the profile likelihood root $r_p(\psi)$, the profile Wald statistic $w_p(\psi)$, the adjusted profile Tsallis root $r_{Sp}(\psi)$ and the profile Tsallis Wald statistic $w_{Sp}(\psi)$, both under the central model and under a contaminated model. 
we note that, under the central model, the two likelihood-based CDs show a reasonably good performance. Apparently, the CD based on the Tsallis $r_p$ is slightly better than its corresponding likelihood-based quantity. This sub-optimal behaviour of $r_p$ under the true model could be explained by the fact that MLE of $\sigma_1,\sigma_2$ is biased in finite samples. Thus, we can conclude that Tsallis, i.e. robust, CDs based on the $r_p$ statistic is a valid alternative to its likelihood-based counterpart. Nevertheless, with contaminated data, only robust $r_p$ CD gives improved inference, i.e. CIs coverage closer to the nominal value, less biased estimators and $p$-values closer to uniform.

%\begin{figure}
%\begin{center}
%\includegraphics[height=6cm,width=8cm]{ICdono}
%\includegraphics[height=6cm,width=8cm]{ICdono1}
%\vspace{-0.3cm}
%\caption{{\small Empirical coverage of bilateral confidence intervals based on several CDs under the central model (right) and under the contaminated model (left), with $\gamma=1.2$.}}
%\label{figsim12}
%\end{center}
%\end{figure}

%Figure~\ref{figsim22} reports the boxplots of themedians of the likelihood and Tsallis based CDs, both under the central model and under a contaminated model. We note that, under the central model, the two estimators present a similar behaviour. On the contrary, under the contaminated model, only the Tsallis estimator present a robust performance. 

%
%\begin{figure}
%\begin{center}
%\includegraphics[height=6cm,width=8cm]{Tsallispoint}
%\includegraphics[height=6cm,width=8cm]{tsallispointcont}
%\vspace{-0.3cm}
%\caption{{\small Likelihood and Tsallis point estimates under the central model (right) and under the contaminated model (left), with $\psi=2$ and $\gamma=1.2$.}}
%\label{figsim22}
%\end{center}
%\end{figure}

%Finally, Figures~\ref{figsim32} and ~\ref{figsim42} report the uniform quantile-quantile plots of the p-values from the four CDs when testing $H_0: \psi=\psi_0$ against $H_1: \psi>\psi_0$, under the central model and under the contaminated model. We note that, under the central model, all the CDs present a reasonable performance. On the contrary, under the contaminated model, the robust CD (\ref{cd2}) presents the better performance. 

%\begin{figure}
%\begin{center}
%\includegraphics[height=6cm,width=8cm]{pvaluestsallisNC}
%\vspace{-0.3cm}
%\caption{{\small Likelihood and Tsallis p-values under the central model, with $\gamma=1.2$.}}
%\label{figsim32}
%\end{center}
%\end{figure}
%
%\begin{figure}
%\begin{center}
%\includegraphics[height=6cm,width=8cm]{pvalttestcont}
%\vspace{-0.3cm}
%\caption{{\small Likelihood and Tsallis p-values under the contaminated model, with  $\gamma=1.2$.}}
%\label{figsim42}
%\end{center}
%\end{figure}

\vspace{0.2cm}

\noindent {\bf Case study.} The debate whether statins can have  adverse effects on cognitive decline in elderly been raging since their introduction in 1987. The dataset considered here (see Mandas  {\em et al.}, 2014) contains measurements on the Mini Mental Score (MMSE) on 329 subjects aged  65 or older, living in a little Sardinia village, collected in 2014. Two groups of subjects are considered: 59 cases treated with statins (average and sample standard deviation MMSE 24.76 and 3.47, respectively) and 270 controls (average and sample standard deviation 26.05 and 3.41, respectively). The left plot of Figure \ref{figfim} illustrates the boxplots of the MMSE in the two groups. Due to the presence of several outliers, the normal assumption for the control may be questionable. To highlight the impact of the outlying observations, we consider a "cleaned" version of the dataset by eliminating three control patients with smallest MMSE.

\begin{figure}
\begin{center}
\includegraphics[height=6cm,width=6cm]{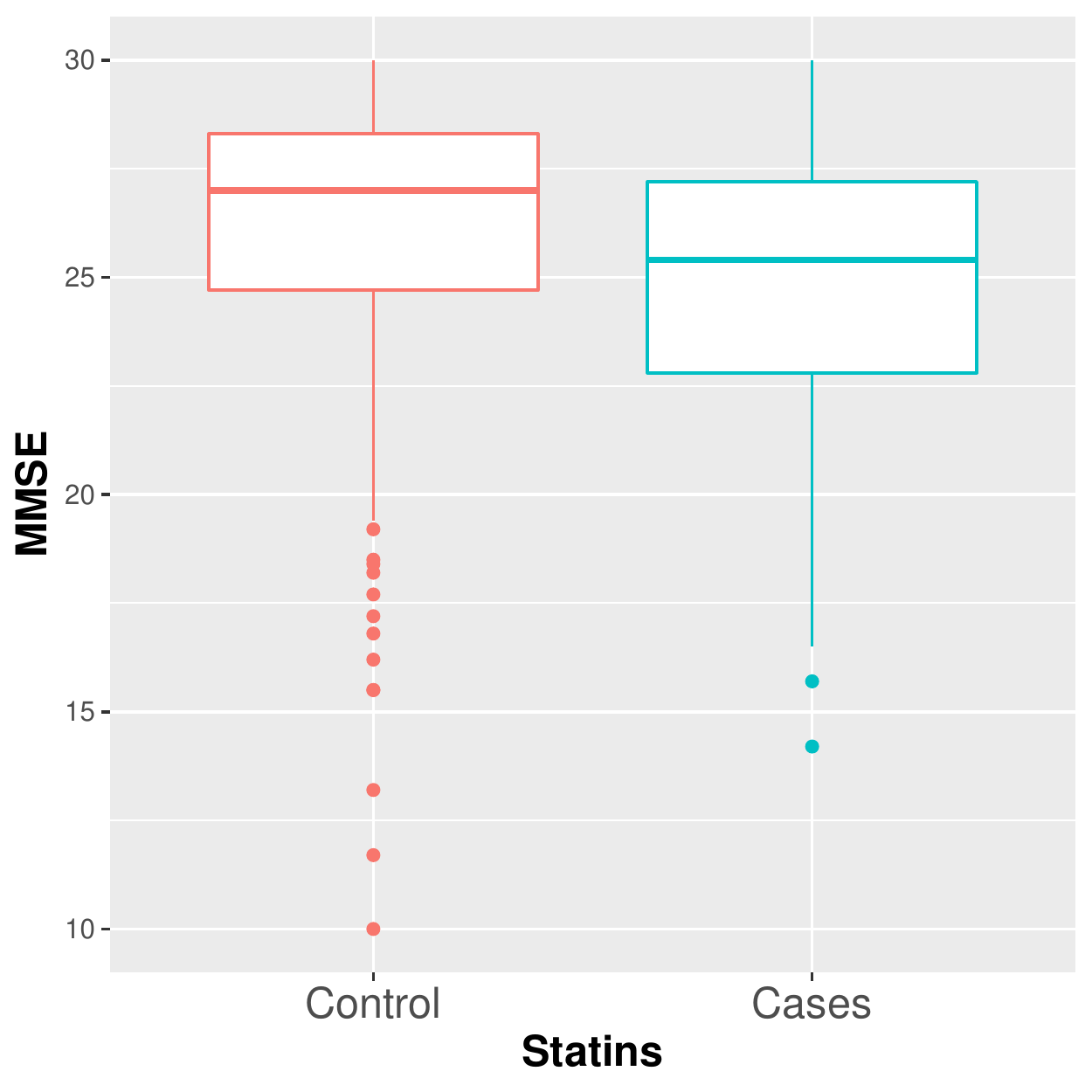}
\includegraphics[height=6cm,width=6cm]{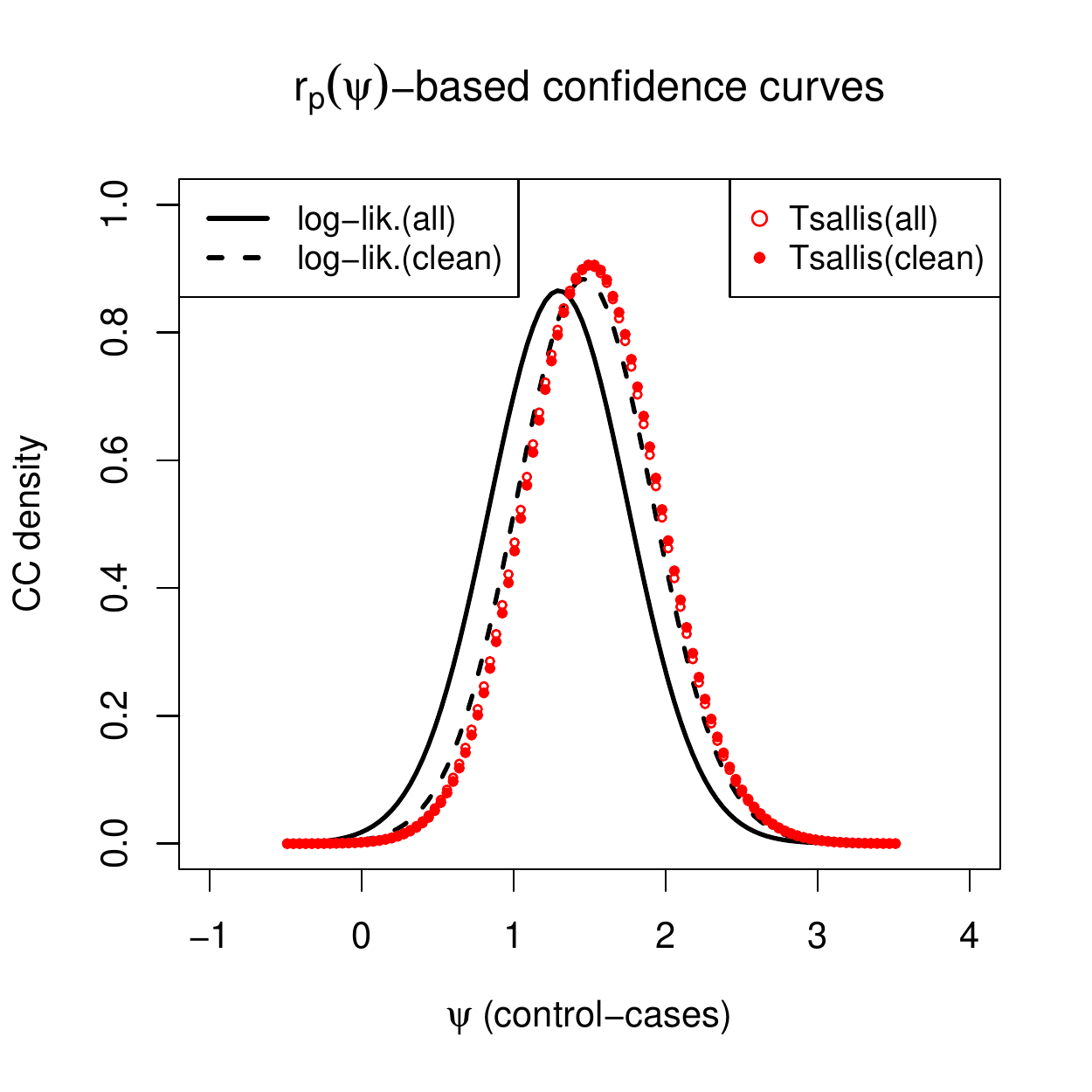}
\vspace{-0.3cm}
\caption{{\small Left: boxplot of MMSE for case and control patients. Right: confidence curves based the pivot $r_p(\psi)$ obtained from the likelihood and the Tsallis scoring rule, using either complete (all) data or "cleaned" data.}}
\label{figfim}
\end{center}
\end{figure}

In this application interest is on the group mean difference, i.e.\ for cases minus controls and the aim is to test for the efficacy of the treatment by testing $H_0 : \psi = 0 $ against $H_1: \psi \neq 0$, or by computing a measure of evidence for $\psi<0$. From the right plot of Figure \ref{figfim} %shows the CCs based on $r_p(\psi)$ and on $r_{Sp}(\psi)$ using both the complete dataset and a modified version obtained by removing three most outlying observations in the control group . Clearly 
we notice that the three outlying control observations have a substantial impact on the likelihood-based CC. On the other hand, the CC based on the Tsallis scoring rule with or without the outlying observations remains essentially unchanged.

Table~\ref{fimt} gives the $p$-value for testing $H_0 : \psi = 0 $ against $H_0: \psi \neq 0$, the measure of evidence for $\psi<0$ and the median of the CCs, with the complete and the cleaned datasets. Note that the CDs based on the complete data and on the Tsallis scoring rule have similar summaries. On the other hand, the CD based on the log-likelihood with the elimination of the outliers gives quite different evidences.

%\begin{figure}
%\begin{center}
%\includegraphics[height=5cm,width=7cm]{CDMM}
%\vspace{-0.3cm}
%\caption{{\small CDs based on $r_p(\psi)$ and on $r_{Sp}(\psi)$ both with the complete data and with the elimination of the outliers in the control group.}}
%\label{figfim2}
%\end{center}
%\end{figure}

\begin{table}
\begin{center}
\begin{tabular}{|r|c|c|c|c|} \hline
Data                     & $r_p(\psi)$       &  $p$-value                  &  Evidence for $\psi<0$  & Median \\ \hline
complete             & \multirow{2}{*}{log-likelihood}     &  $5.7\times10^{-3}$  &  $2.9\times 10^{-3}$ & -1.30 \\
cleaned               &     & $1.5\times 10^{-3}$   &  $7.7\times 10^{-4}$ & -1.47 \\ \hline
complete             &  \multirow{2}{*}{Tsallis}              &  $5.5\times 10^{-4}$  &  $2.8\times 10^{-4}$ &-1.51 \\
cleaned                &               &  $4.9\times 10^{-4}$  &  $2.5\times 10^{-4}$ &-1.52 \\ \hline
\end{tabular}
\vspace{0cm}
\caption{{\small Summaries of CCs based on the MMSE data obtained from the $r_p(\psi)$ pivot.}}
\label{fimt}
\end{center}
\end{table}

%\begin{table}
%\begin{center}
%\begin{tabular}{|c|c|c|c|c|} \hline
%data                       & pivot  &  $CC(0)$ &  p-value & median \\ \hline
%(diff1,diff2)             & $r_p(\psi)$        &  0.027 &   0.058 & -1.09 \\
%(diff1,diff2$>15$)   & $r_p(\psi)$        & 0.001  &  0.002 & -1.71 \\ \hline
%(diff1,diff2)             & $r_{Sp}(\psi)$   & 0.021 &   0.055 & -1.05 \\
%(diff1,diff2$>15$)   & $r_{Sp}(\psi)$   & 0.020 &   0.041 & -1.36 \\ \hline
%\end{tabular}
%\vspace{-0.3cm}
%\caption{{\small CCs summaries based on $r_p(\psi)$ and on $r_{Sp}(\psi)$ both with the complete data and with the elimination of the outliers in the case group.}}
%\label{fimt}
%\end{center}
%\end{table}

%%%%%%%%%%%%%%%%%%%%%%%%%%%%%%%%%%%%%%%%%
%%%%%%%%%%%%%%%%%%%%%%%%%%%%%%%%%%%%%%%%%

\subsection{Area under the ROC curve}

Let $X_1$ and $X_2$ be independent random variables with distributions $F_{X_1} (x_1; \theta_1 )$ and $ F_{X_2}(x_2;\theta_2)$, respectively. A stress-strength model is concerned with the problem of evaluating $P ( X_1 < X_2 )$. For instance, in a clinical study, $X_1$ may be the response of a control group, $X_2$ the response of a treatment group and the reliability parameter $P(X_1<X_2)$ measures the effectiveness of the treatment as given by the area under the ROC curve.

By the definition of reliability, $P(X_1<X_2)$ can be evaluated as a function of the parameter $\theta = (\theta_1 , \theta_2 )$, through the relation
\[
\psi = \psi(\theta) = P(X_1 < X_2) = \int F_{X_1} (t;\theta_1) \, d F_{X_2} (t;\theta_2).
\]

Theoretical expressions for $\psi$ are available under several distributional assumptions both for $X_1$ and $X_2$ (see Kotz {\em et al.}, 2003). For instance, if $X_1$ and $X_2$ are independent normal random variables, i.e. $X_1 \sim N(\mu_1,\sigma^2_1)$ and $X_2 \sim N(\mu_2,\sigma^2_2)$, the reliability parameter is $\psi = \Phi \left( \frac{\mu_2 - \mu_1}{\sqrt{\sigma^2_1+\sigma^2_2}} \right)$. 

\vspace{0.2cm}

\noindent {\bf Simulation results.}  Let $(x_{11},\ldots,x_{1n_1})$ and $(x_{21},\ldots,x_{2n_2})$ be independent samples from two exponential distributions with parameters $\lambda_1$ and $\lambda_2$. Since $E(X_1)= 1/\lambda_1$ and $E(X_2) = 1/\lambda_2$, then $\psi = \frac{\lambda_1}{\lambda_1+\lambda_2}$, which is the parameter of interest.

In this situation the total Tsallis score is 
%\[
%S(\lambda_1,\lambda_2) = \frac{n_1\lambda_1^{\gamma}+n_2\lambda_2^{\gamma}}{1+\alpha} - \left(1+\frac{1}{\gamma}\right)\left(\lambda_1^{\gamma}\sum_{j=1}^{n_1} e^{-\gamma \lambda_1x_{1j}} + \lambda_2^{\gamma}\sum_{j=1}^{n_2} e^{-\gamma \lambda_2 x_{2j}}\right).
%\]
\[
S(\lambda_1,\lambda_2) = \left(1-\frac{1}{\gamma}\right)\left(n_1\lambda_1^{\gamma-1}+n_2\lambda_2^{\gamma-1}\right)-\gamma\left[\lambda_1^{\gamma-1}\sum_{j=1}^{n_1} e^{-(\gamma-1) \lambda_1x_{1j}} +\lambda_2^{\gamma-1}\sum_{j=1}^{n_2} e^{-(\gamma-1) \lambda_2 x_{2j}}\right].
\]
In the simulations, we set the true AUC to $\psi = 0.85$ and the nuisance parameter $\lambda_2 = 2/3$; thus the implied true rates for the exponential distributions are $\lambda_1=3.778$  and $\lambda_2 = 2/3$. The robustness tuning parameter $\gamma$ is fixed in such a way that the resulting estimator is 10\% less efficient than the MLE, under the true model. We generated $10^5$ datasets with sizes of the two samples $(n_1=20,n_2=40)$ and computed the coverage of Wald- and $r_p(\psi)$-type confidence intervals at various confidence levels. Furthermore, the uniformity of the $p$-values when testing $H_0:\psi=\psi_0$ against $H_1:\psi<\psi_0$ is also checked. Data are generated from the true model, with and without contamination. To generate a contaminated dataset, the last observation of the first sample, i.e. the one with $n_1$ observations, is shifted by adding 3. The Wald-type confidence intervals and the associated $p$-values were computed with $\psi$ reparametrized on the logit scale; the confidence intervals were then re-transformed back to the original $[0,1]$ scale.

\begin{figure}
\includegraphics[width=0.44\textwidth]{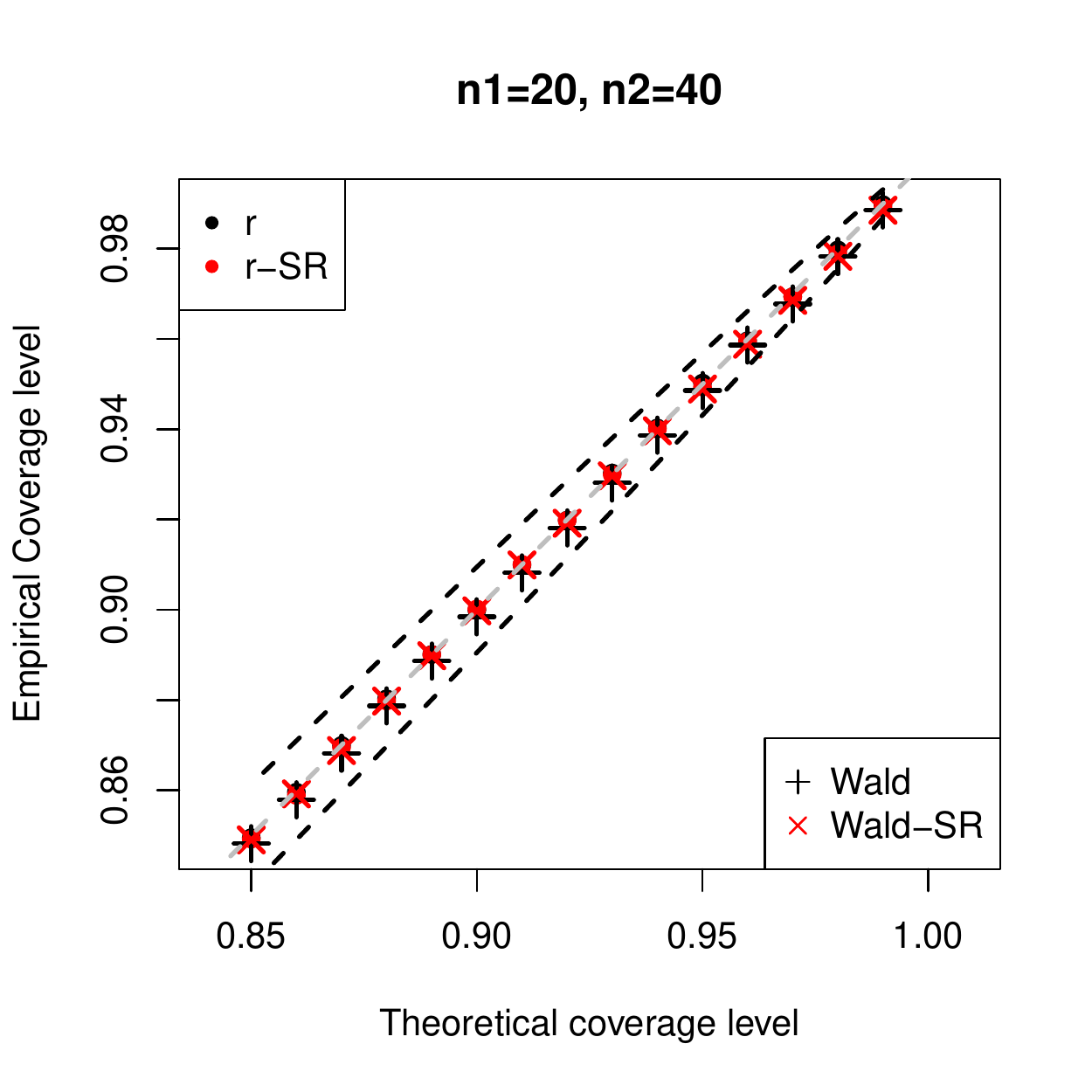}
\includegraphics[width=0.44\textwidth]{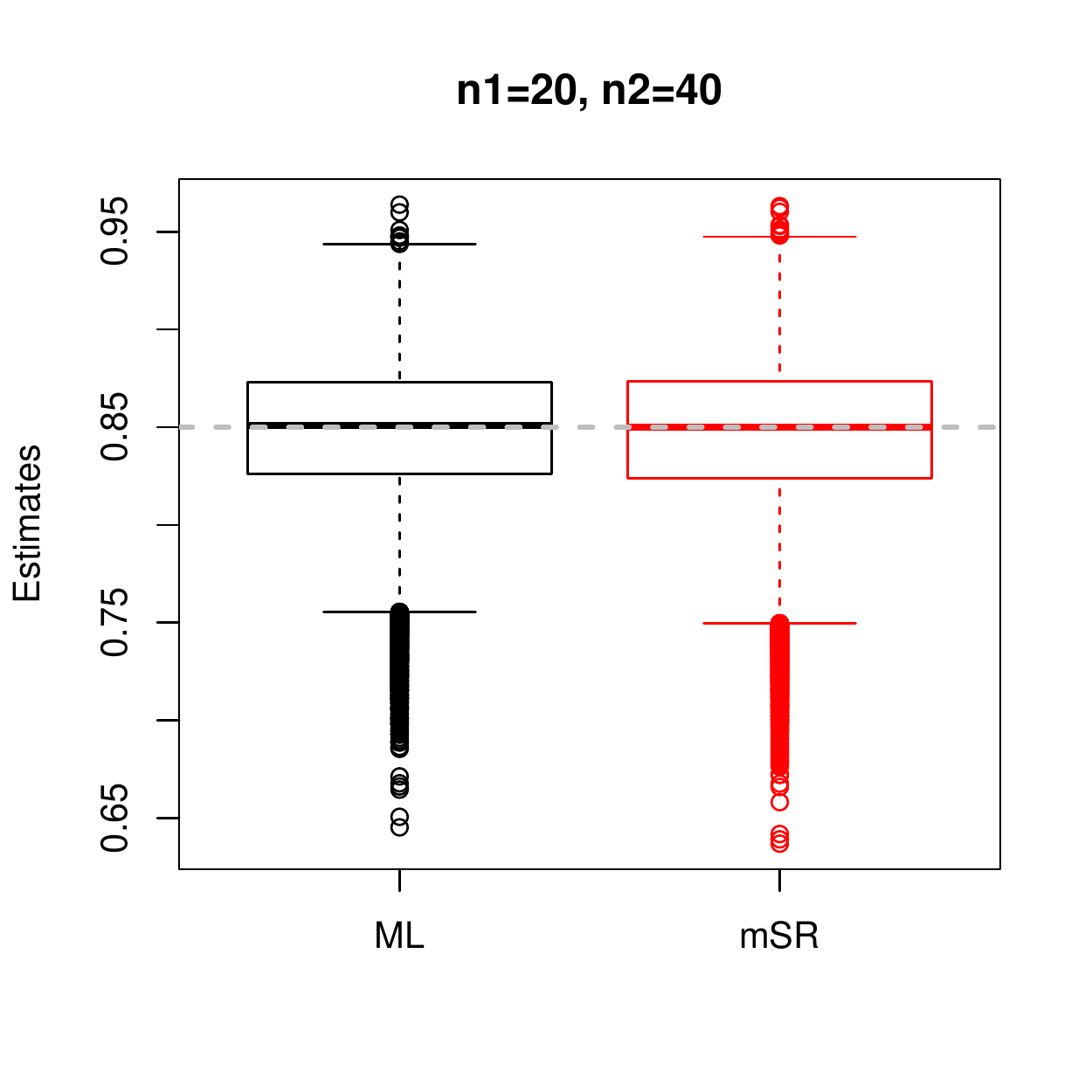}\\
\includegraphics[width=0.44\textwidth]{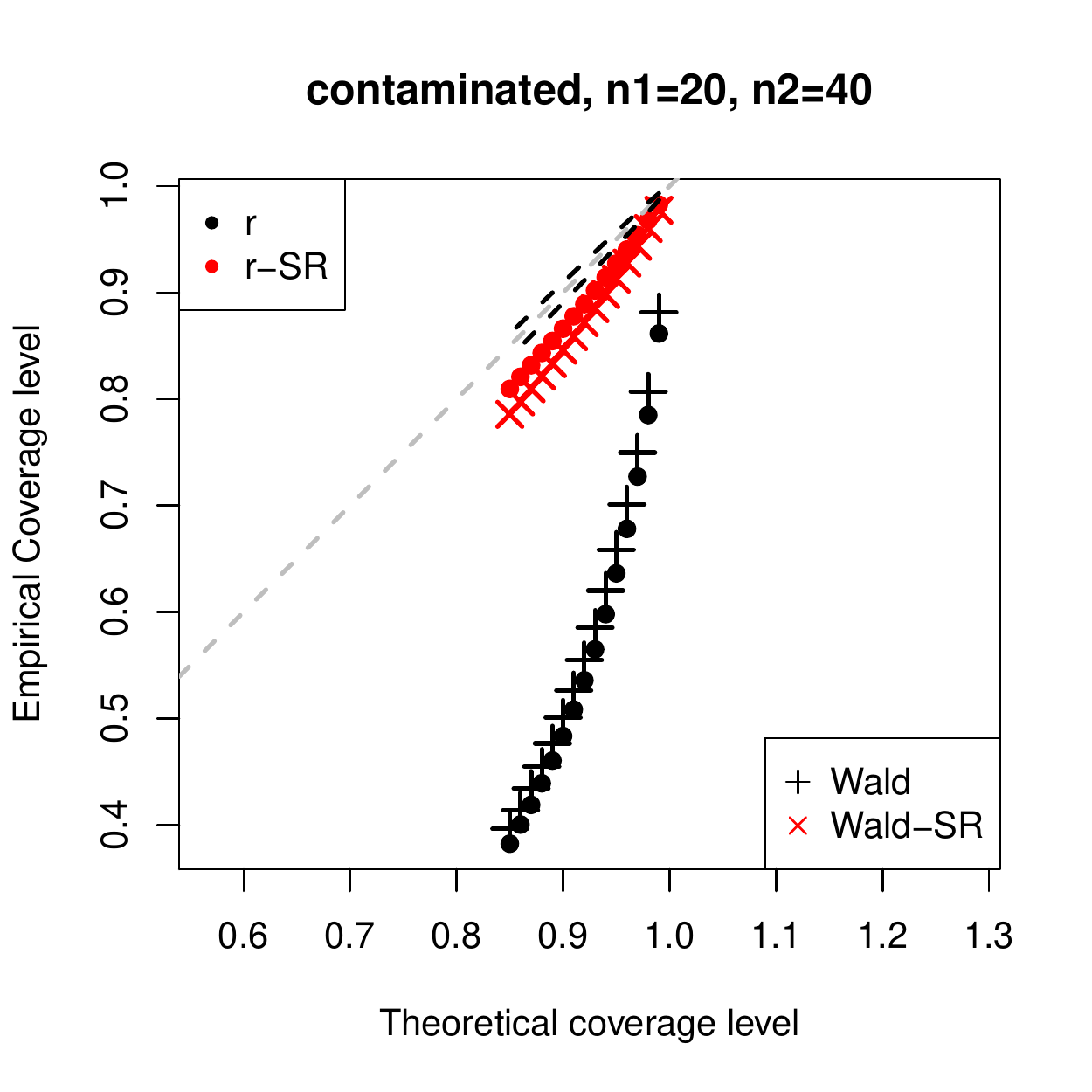}
\includegraphics[width=0.44\textwidth]{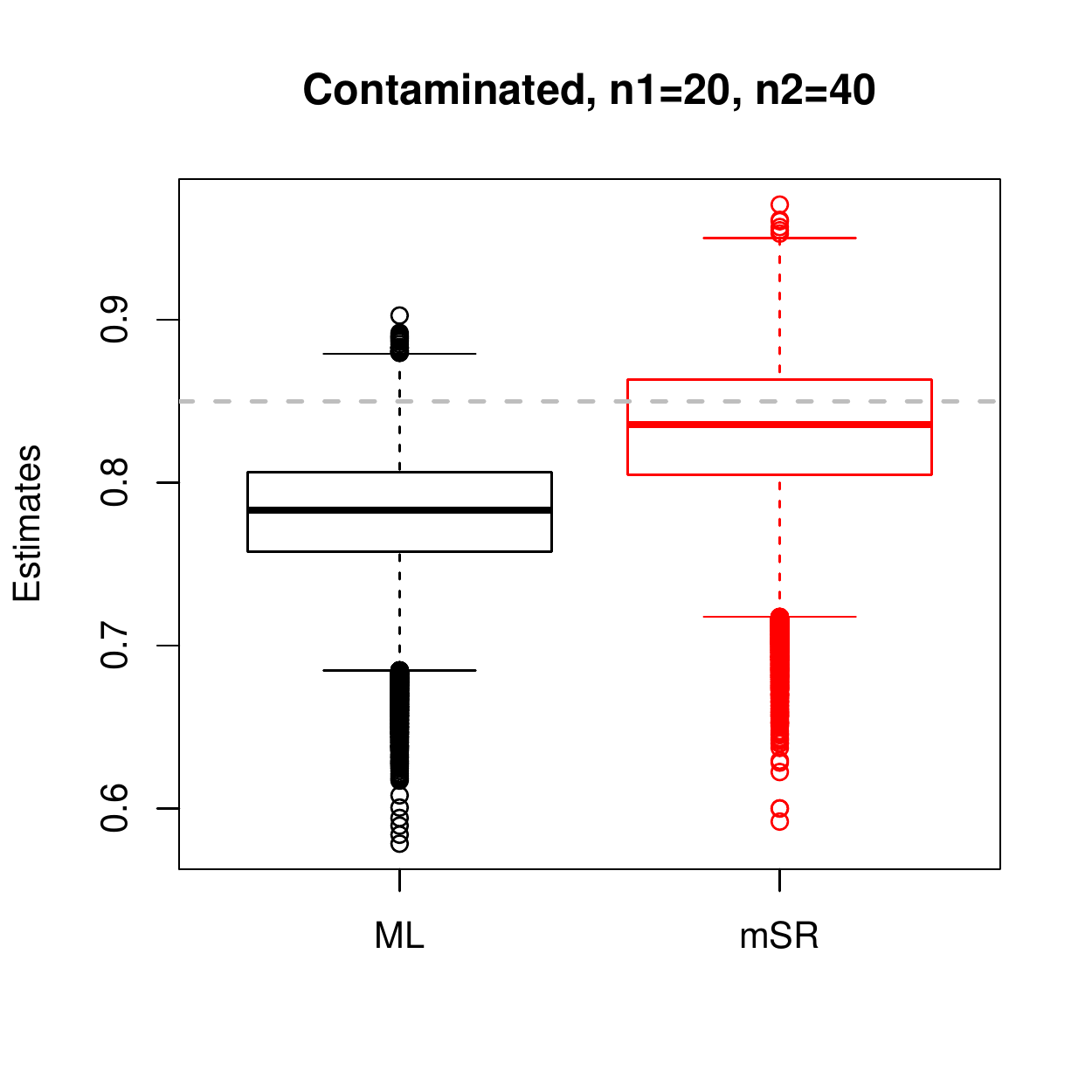}\\
\caption{\small AUC for exponential random variables. Empirical coverage of CIs and distribution of the CD estimates with data generated under the true model without contamination (first row) and with contamination (second row). Dashed lines represent 10 $\times$ Monte Carlo standard error from the theoretical confidence level.}\label{fig:auc_exp}
\end{figure}

\begin{figure}
\includegraphics[width=0.24\textwidth]{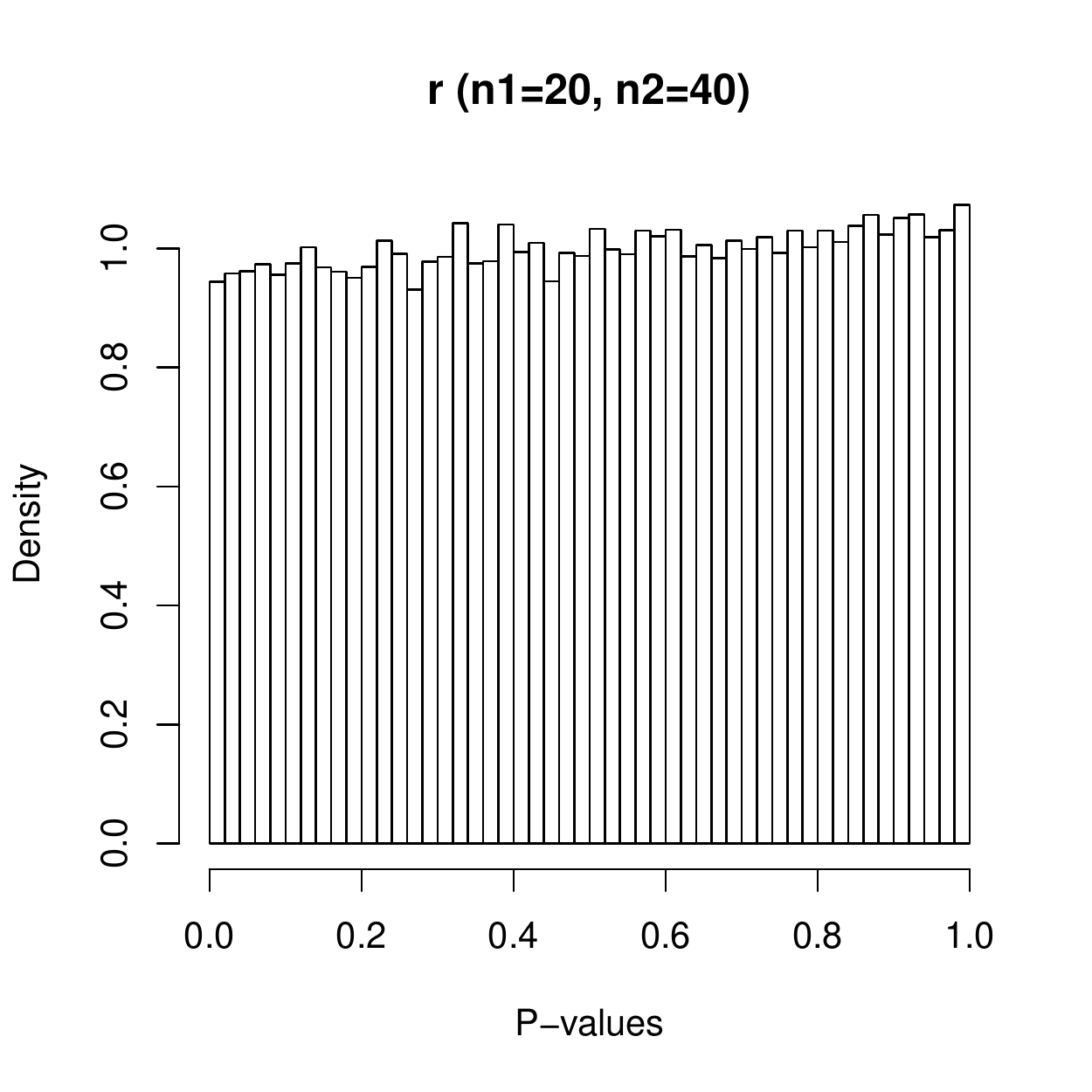}
\includegraphics[width=0.24\textwidth]{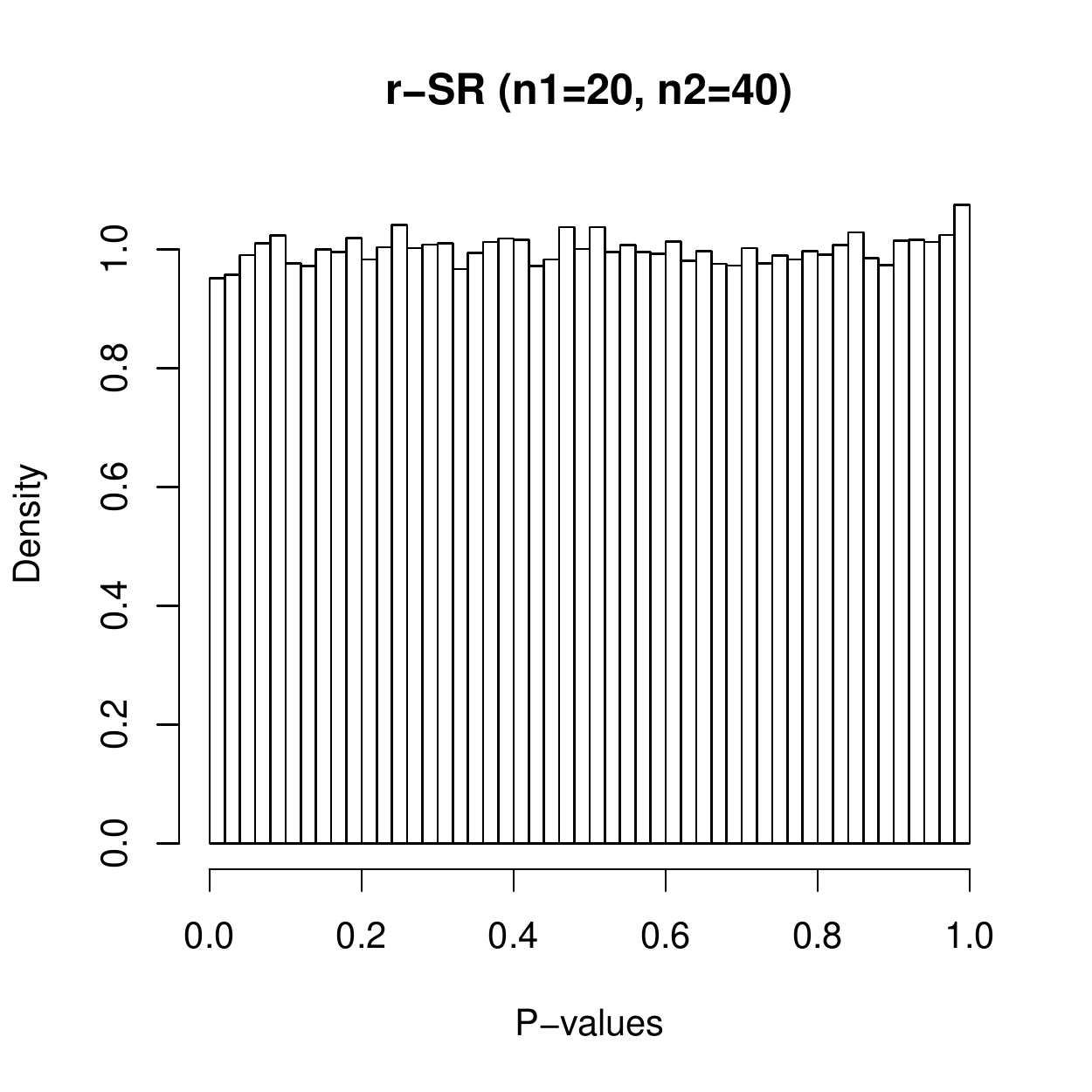}
\includegraphics[width=0.24\textwidth]{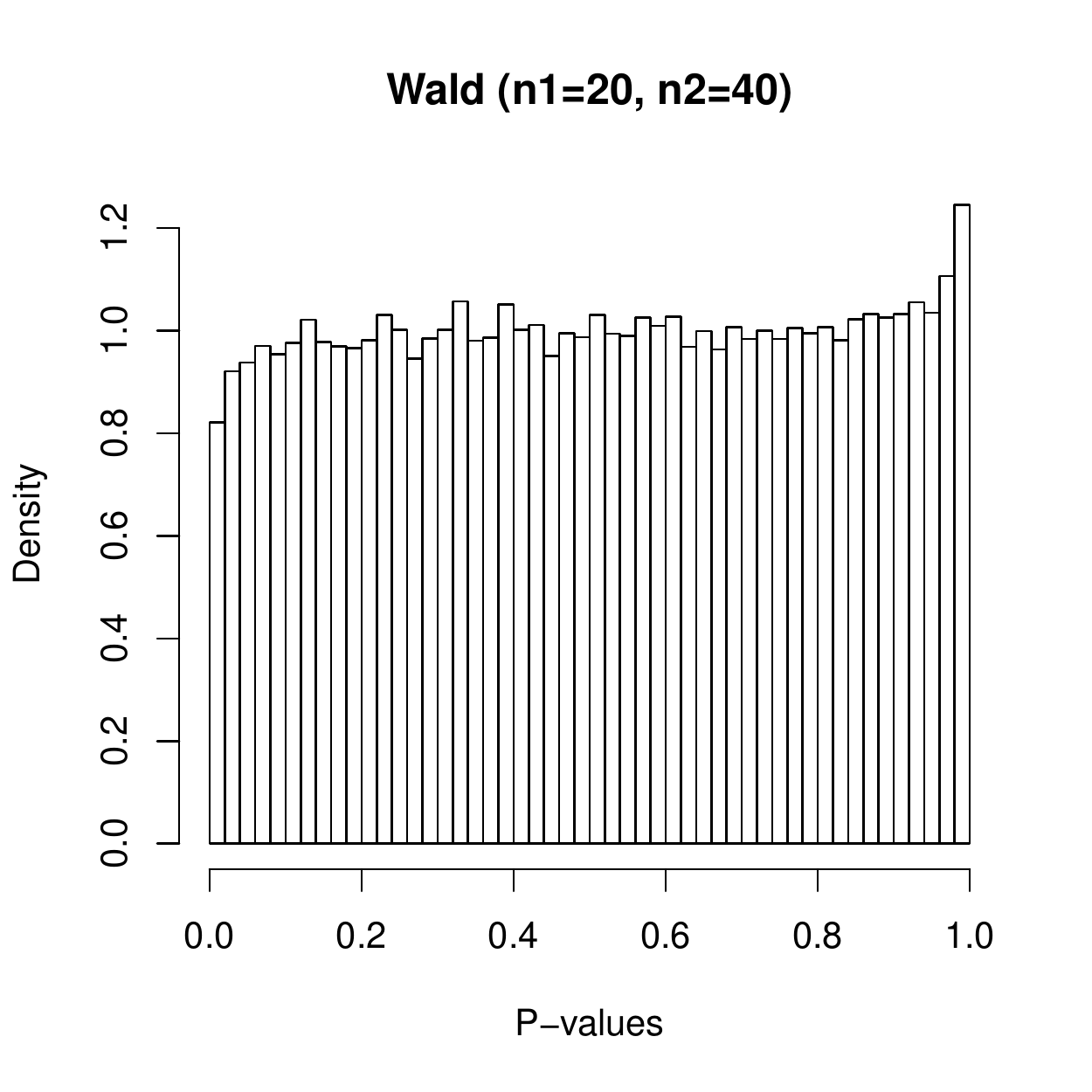}
\includegraphics[width=0.24\textwidth]{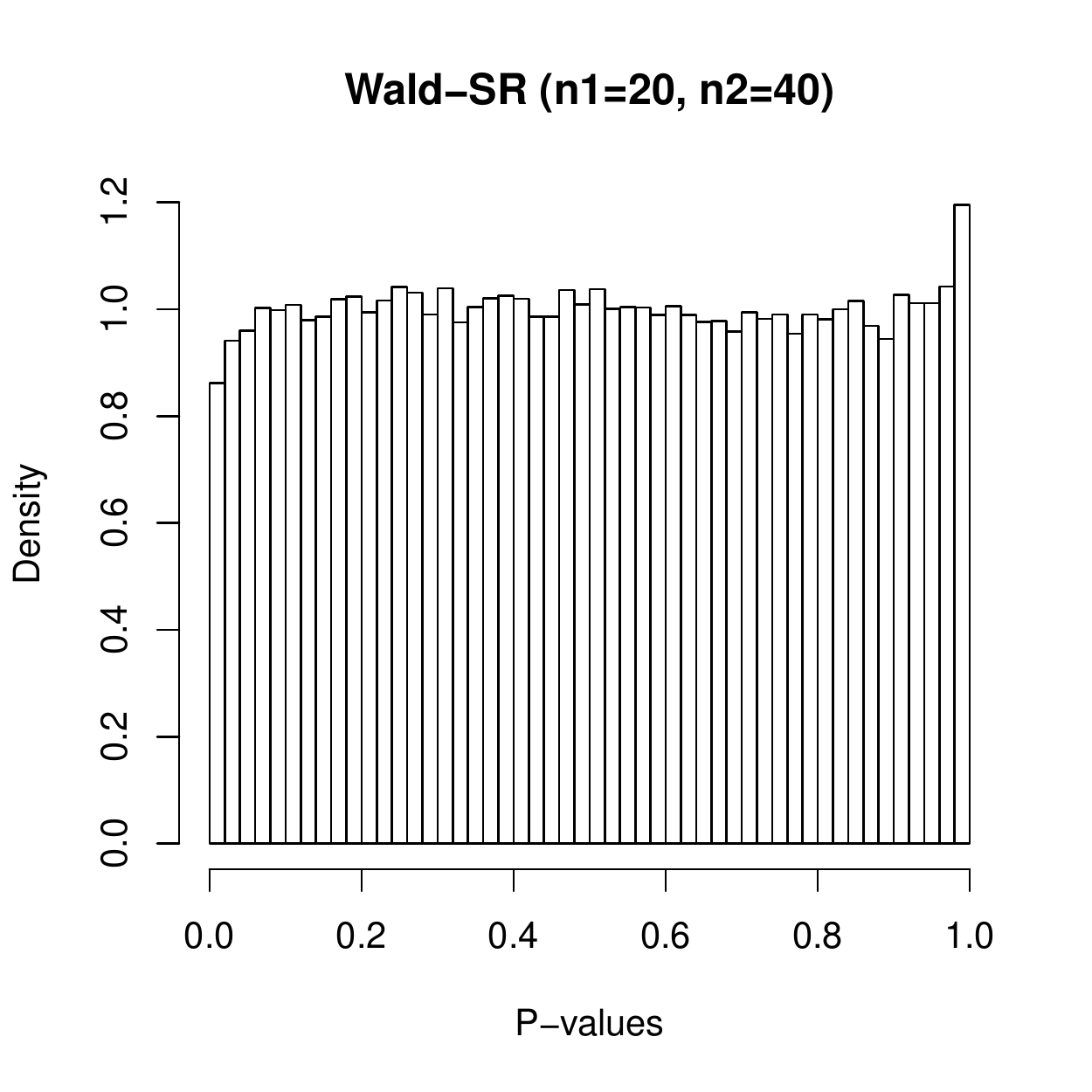}
\includegraphics[width=0.24\textwidth]{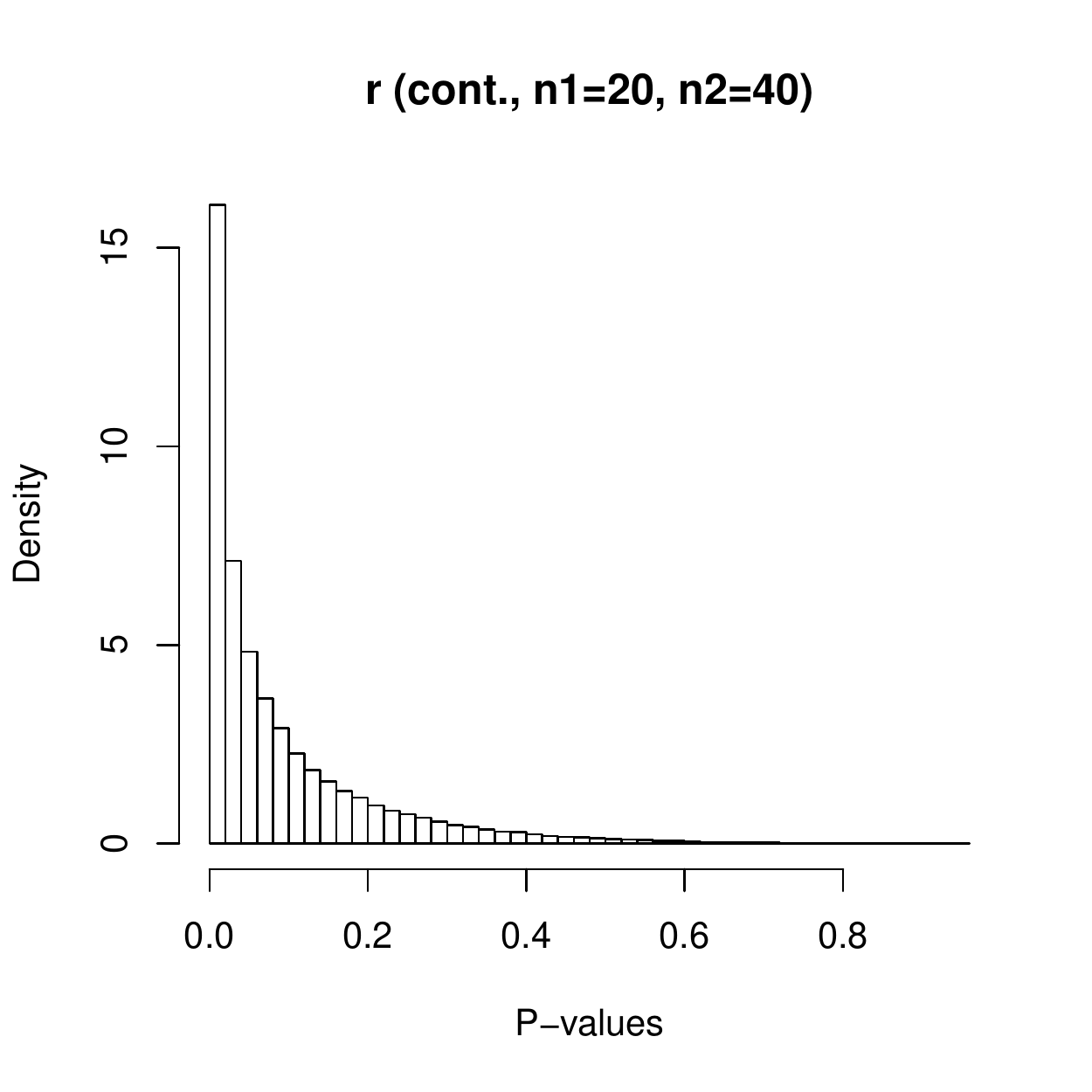}
\includegraphics[width=0.24\textwidth]{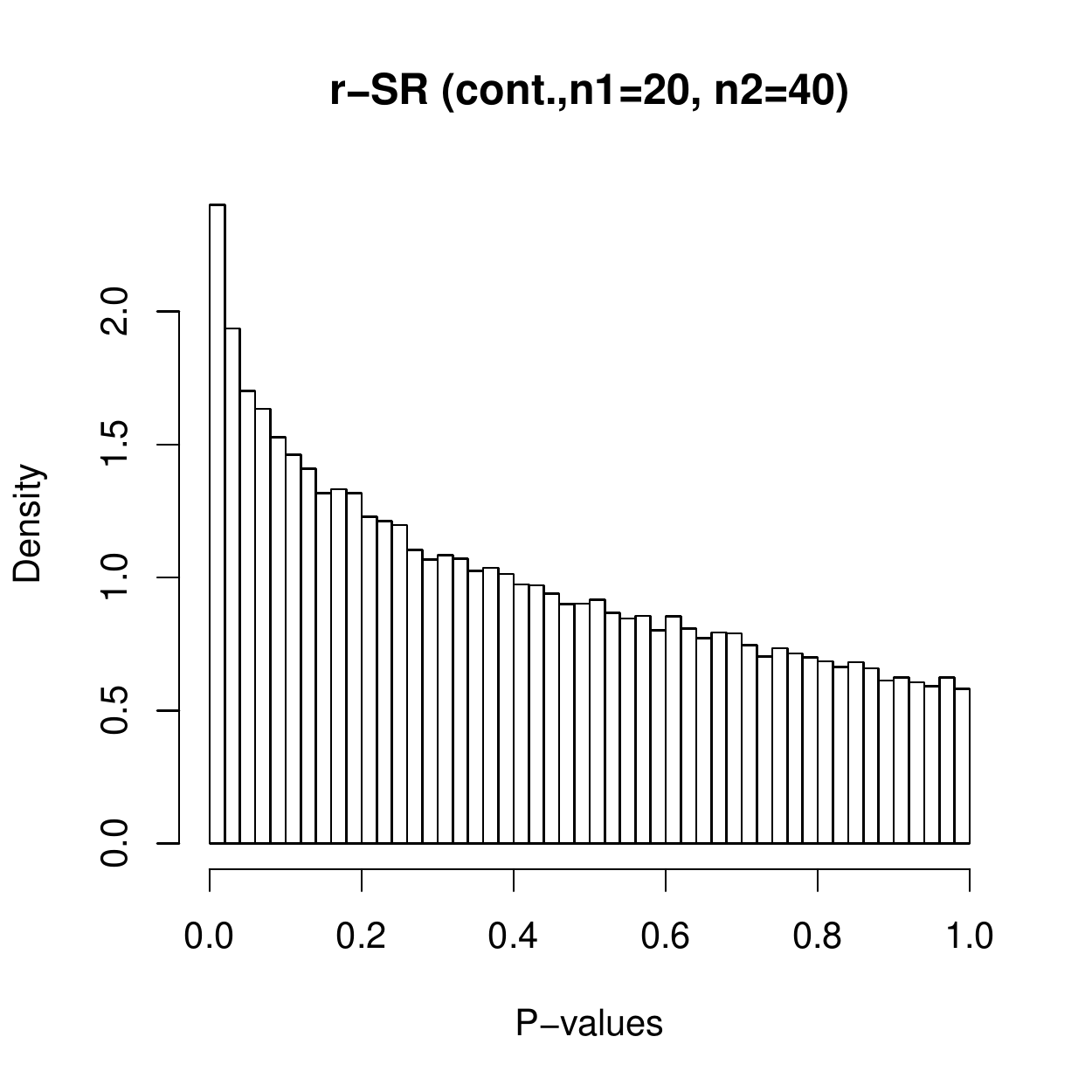}
\includegraphics[width=0.24\textwidth]{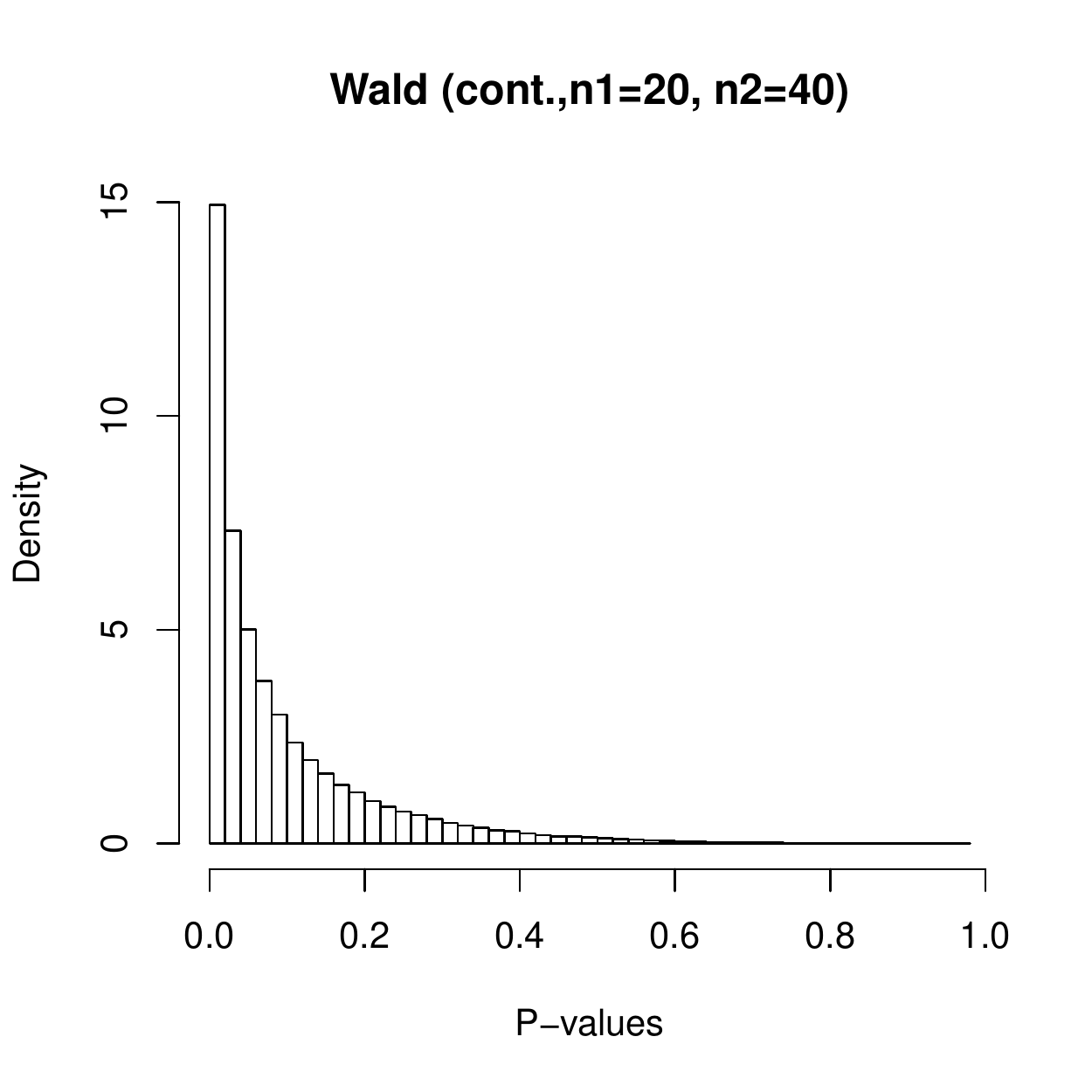}
\includegraphics[width=0.24\textwidth]{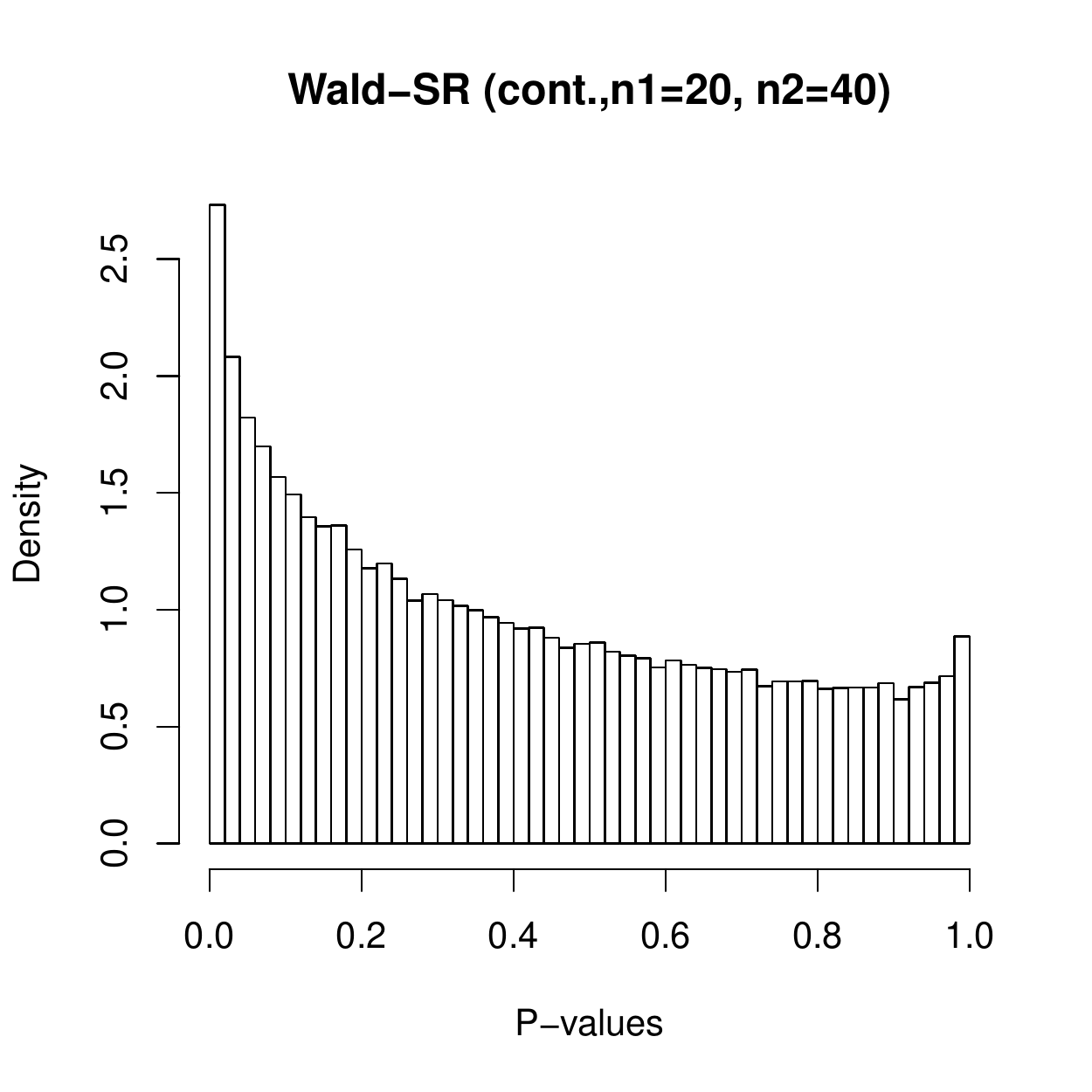}
\caption{\small AUC for exponential random variables. Empirical distribution of p-values for $H_0:\psi = 0.85$ against $H_1:\psi < 0.85$. Figures on the first row are based on non contaminated data, those of the second row are based on contaminated data.}\label{fig:auc_exp_pval}
\end{figure}

%\textcolor{red}{The results, illustrated in Figures \ref{fig:auc_exp} and \ref{fig:auc_exp_pval}, are those of  $(n_1=5,n_2=10)$ and $(n_1=20,n_2=40)$; results for $(n_1=10,n_2=15)$, which can be found in the Supplementary Material, are similar. }
From the simulation study, illustrated by Figures~\ref{fig:auc_exp} and \ref{fig:auc_exp_pval}, we can conclude the following. With non contaminated data, the performance of the various CD considered is quite similar. Also the distribution of the $p$-values based on $r_p(\psi)$-type pivots  and Wald-type pivots (see Fig.~\ref{fig:auc_exp_pval}) are very similar, with the former being closer to uniform than those based on Wald-type pivots. Under contaminated data, CDs based on $r_p(\psi)$ and Wald pivots perform poorly, whereas their robust counterparts perform substantially better, with the Tsallis $r_p(\psi)$ being preferred over the Tsallis Wald-type pivot.  The formers also lead to $p$-values being closer to uniform than the $p$-values obtained with the likelihood-based pivots.

\vspace{0.2cm}

\noindent {\bf Case study.} The data come from a study which aimed to assess the role of the HSP70 (Heat Shock Protein 70 kilodaltons) protein on the presence of Anaplastic Large Cell Lymphoma (ALCL, see Ventura and Racugno, 2011). Diseased patients seem to have higher HSP70 levels than healthy subjects. Thus, HSP70 protein levels can be studied as a biomarker for detecting early ALCL lymphoma and, therefore, its effectiveness in diagnosing the disease can be evaluated by $\psi = P (X < Y )$. The data at hand consist of a small sample: 10 patients with ALCL (sample mean and sample standard deviation equal 1.437 and 1.549) and 4 healthy subjects (sample mean and sample standard deviation equal 0.235 and 0.151).  Two independent exponential random variables were assumed for the protein level in both groups of patients.  There appears (see the boxplots in Figure~\ref{figfim}) to be a diseased patient with HSP70 level very different from the rest of the sample, thus in the subsequent analyses we will compute CDs for $\psi$ using both the original dataset and the dataset obtained by deleting the outlying diseased patient.

\begin{figure}
\begin{center}
\includegraphics[height=6cm,width=6cm]{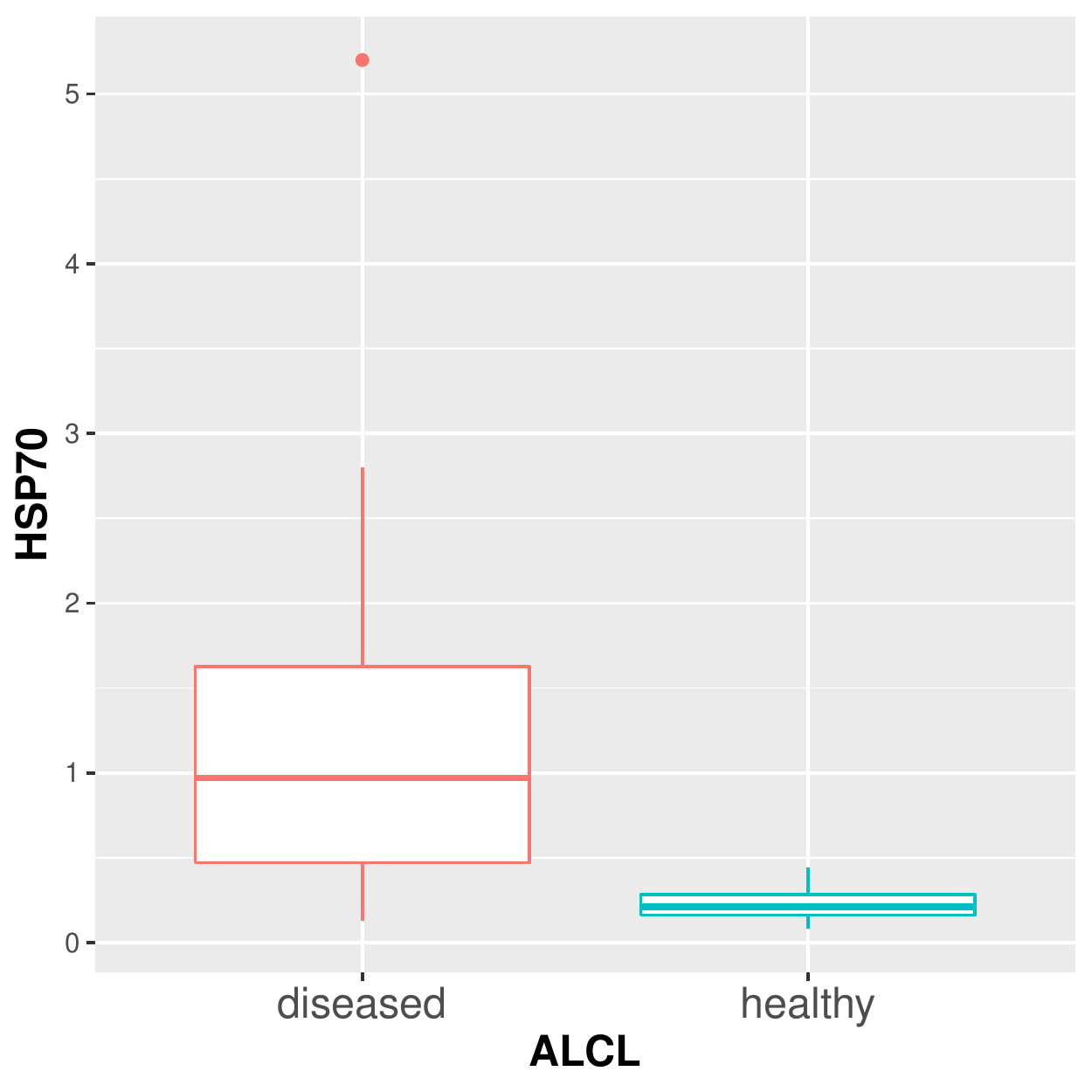}
\includegraphics[height=6cm,width=6cm]{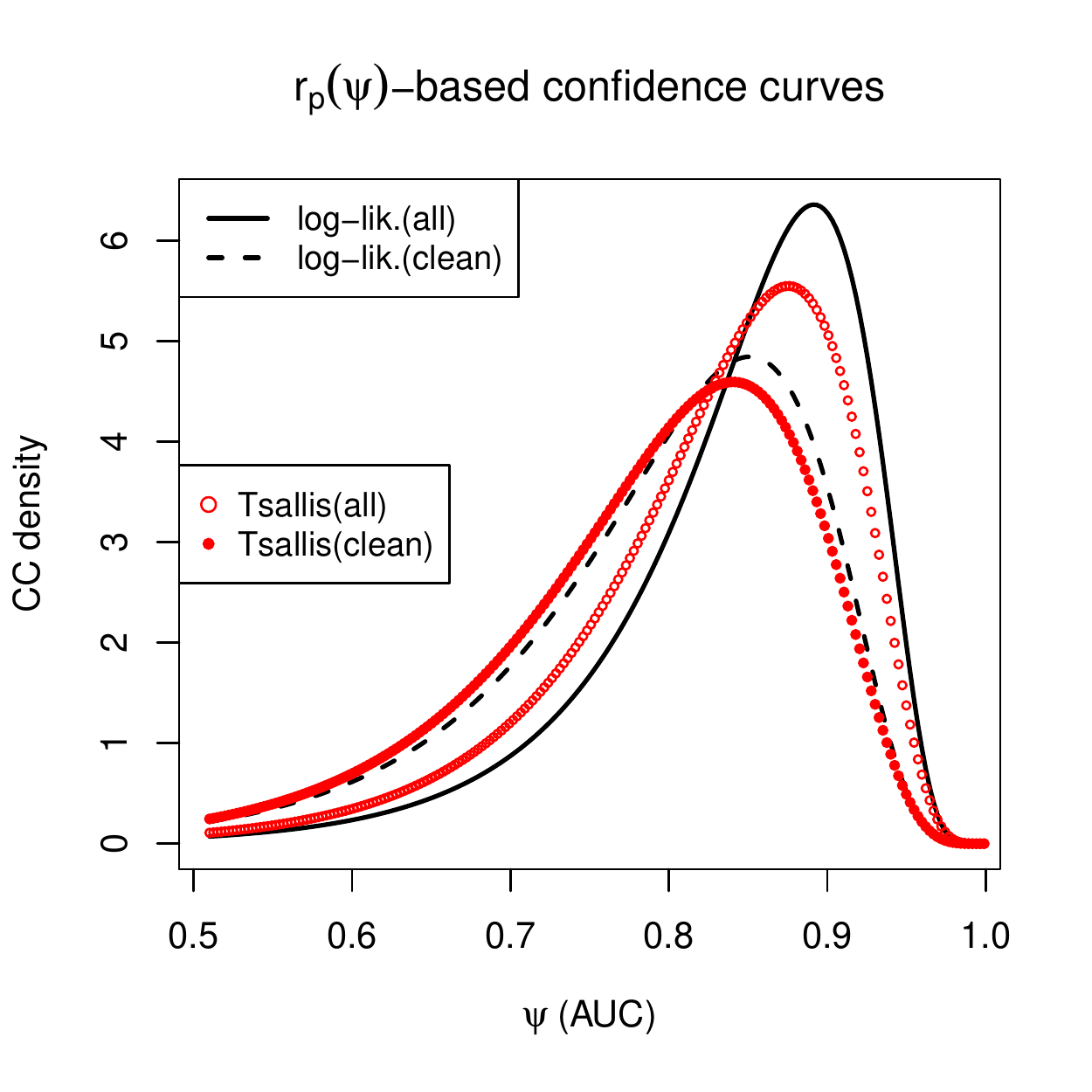}
\vspace{-0.3cm}
\caption{{\small Left: boxplot of HSP70 for case and control patients. Right: confidence curves based the pivot $r_p(\psi)$ obtained from the likelihood and the Tsallis scoring rule, using either complete (all) data or with a "cleaned" version.}}
\label{figfim}
\end{center}
\end{figure}

From the CCs (shown on the right plot of Figure~\ref{figfim}) we note that the CC based on the Tsallis scoring rule is in between the two log-likelihood CCs, which are obtained from the complete and cleaned data, respectively. Furthermore, there also appear a substantial difference between the two Tsallis-based CCs, which is presumably due to the small sample sizes. In particular, the 0.95 CI for the AUC with the Tsallis CCs with complete and cleaned data are (0.596, 0.944) and (0.531 0.929), respectively, whereas those for log-likelihood based are (0.627, 0.948) with the complete data and (0.54, 0.93) with the cleaned data.

%%%%%%%%%%%%%%%%%%%%%%%%%%%%%%%%%%%%%%%%%%
%%%%%%%%%%%%%%%%%%%%%%%%%%%%%%%%%%%%%%%%%%

\subsection{Linear regression model}

Let us consider a linear regression model of the form (\ref{modello}) with $\mu(x_i,\beta)=x_i^{\T} \beta$, $i=1,\ldots,n$. Usually, the parameter of interest is one of the regression coefficients $\beta$.

The Tsallis score is given by (\ref{tnl}). Th Tsallis score estimator is $B$-robust since the influence function is bounded (see Ghosh and Basu, 2013).

\vspace{0.2cm}

\noindent {\bf Simulation results.} For $p=3$, let $\psi=\beta_2$ be the scalar parameter of interest and let $\lambda=(\beta_1,\beta_3,\sigma)$ be the nuisance parameter. As in the previous examples, we ran a simulation experiment with $n=50$ and the robustness constant $\gamma$ is fixed to 1.22.   

\begin{figure}
\begin{center}
\includegraphics[height=5cm,width=7cm]{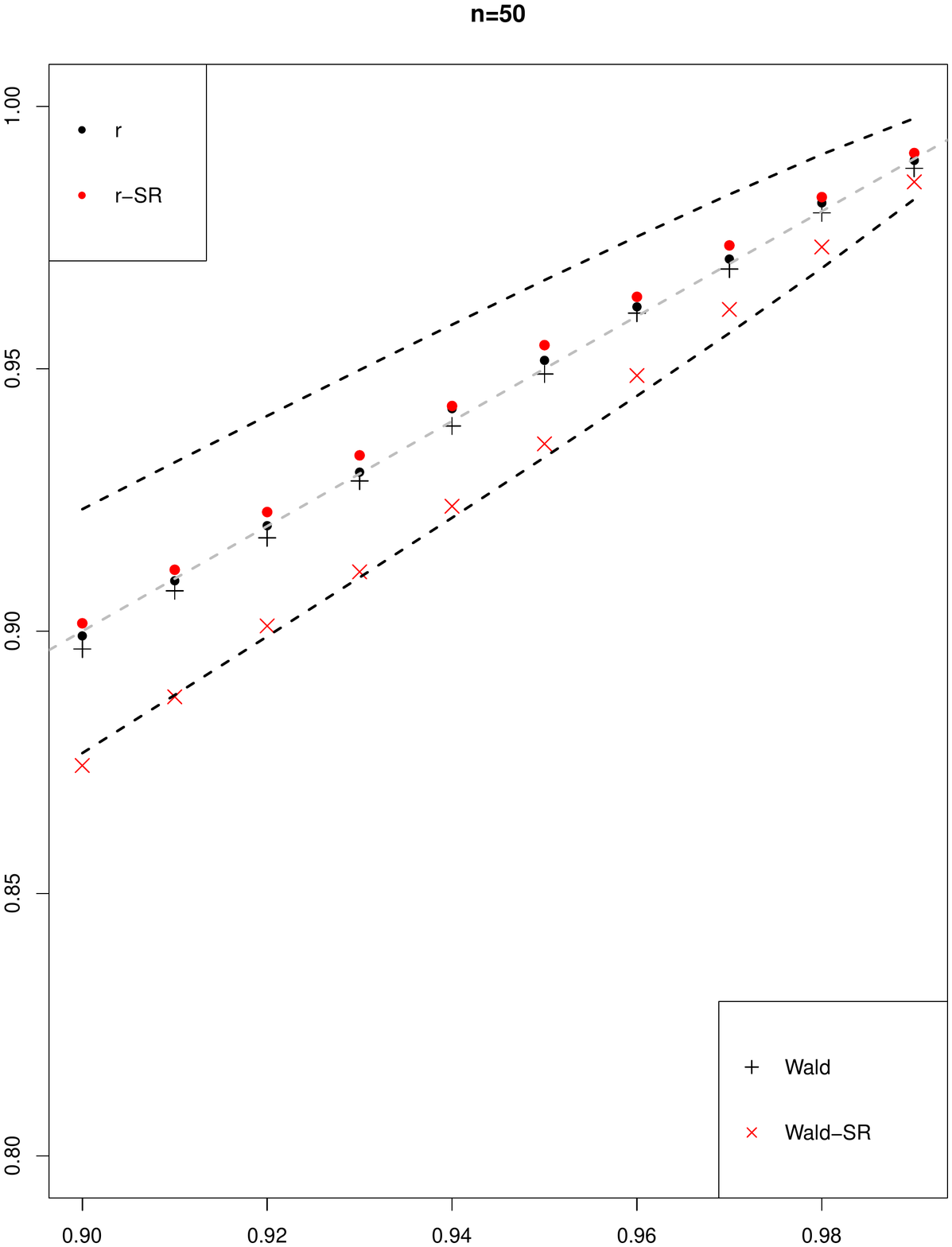}
\includegraphics[height=5cm,width=7cm]{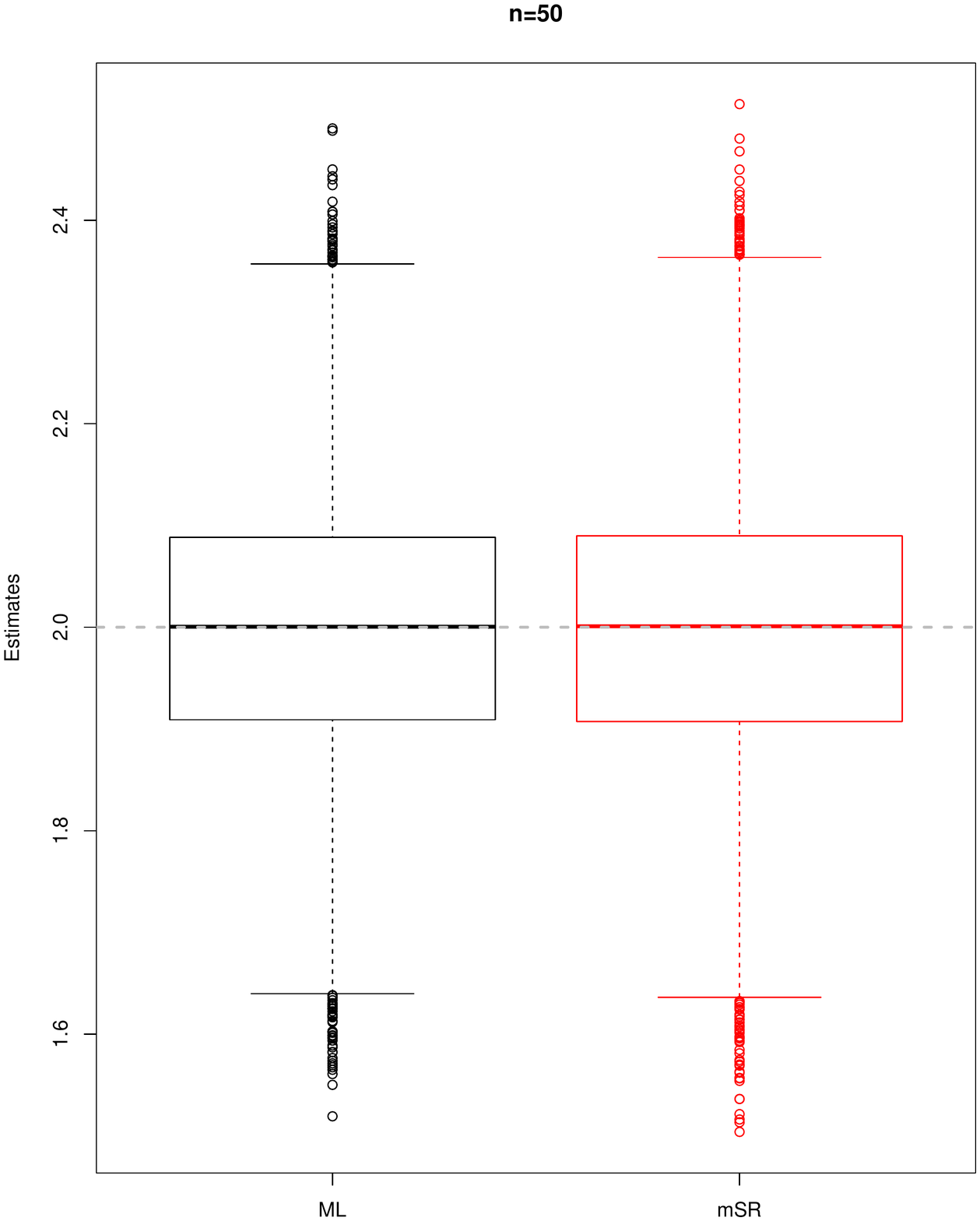}
\includegraphics[height=5cm,width=7cm]{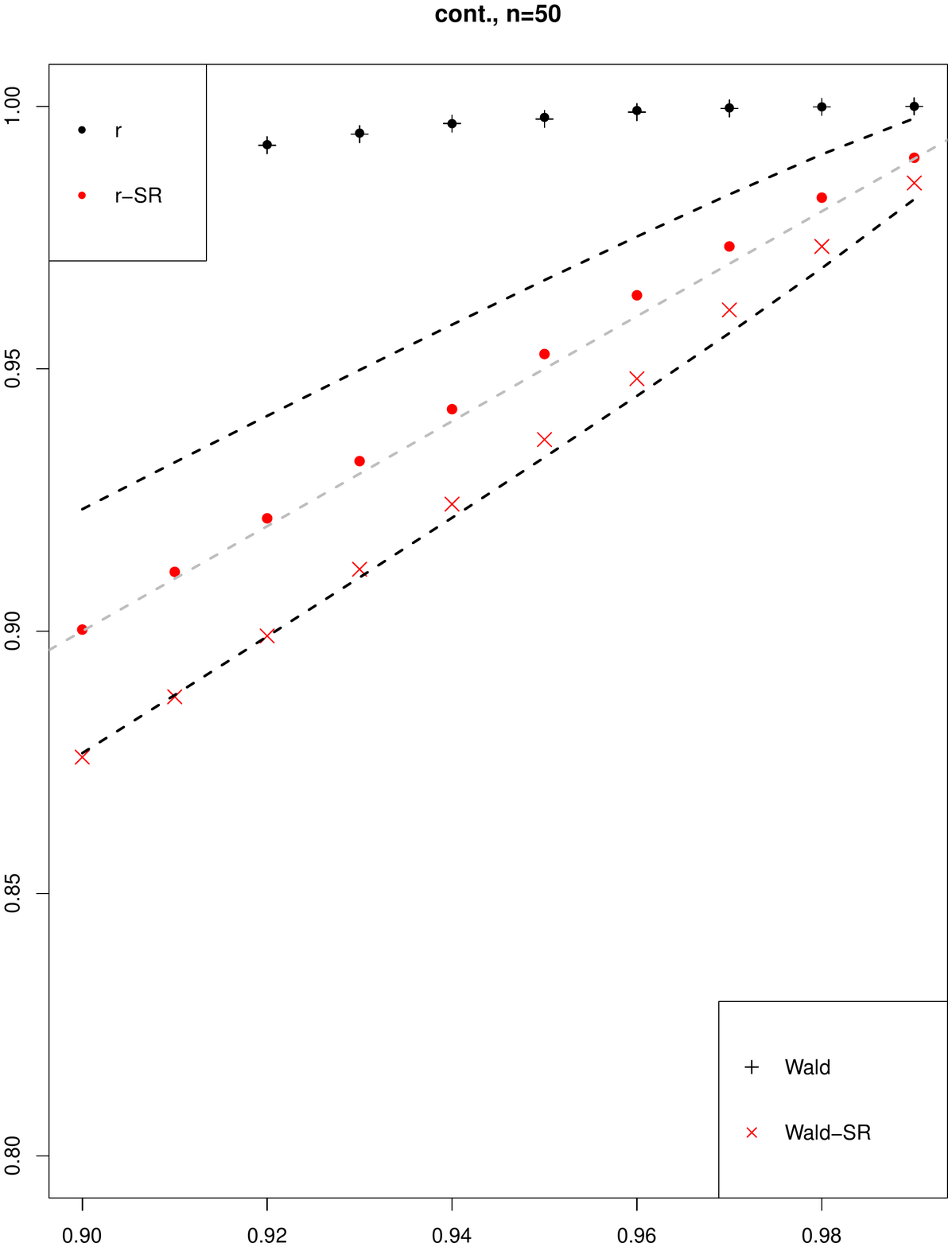}
\includegraphics[height=5cm,width=7cm]{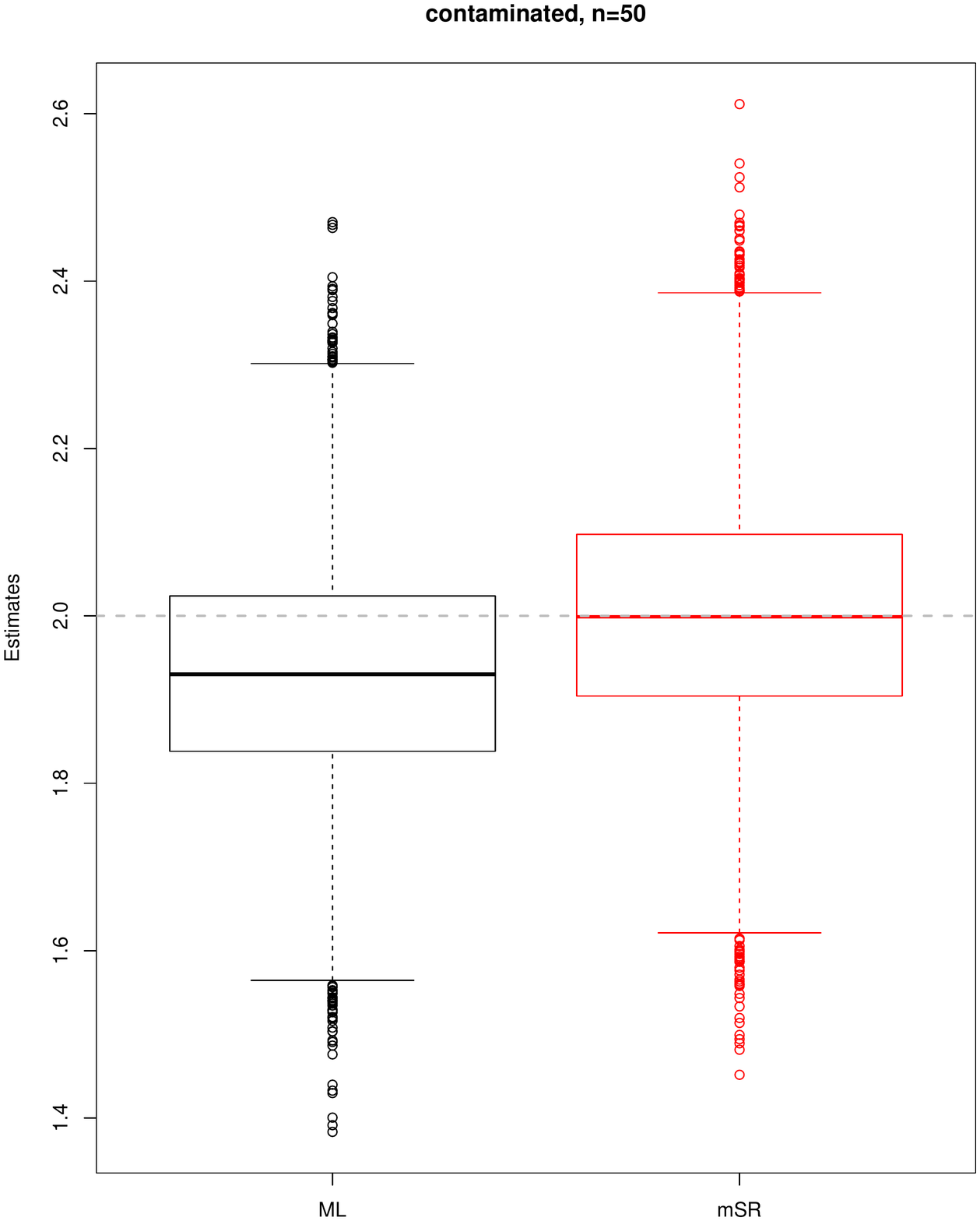}
\vspace{-0.3cm}
\caption{{\small Regression model. Empirical coverage of CIs and distribution of the CD estimates with data generated under the true model without contamination (first row) and with contamination (second row). Dashed lines represent 10$\times$Monte Carlo standard error from the theoretical confidence level.}}
\label{figsim333}
\end{center}
\end{figure}

Figure~\ref{figsim333} (first column) reports the empirical coverages of bilateral confidence intervals based on the four considered CDs, both under the central model and under a contaminated model. The contamination is obtained with shift contamination. As in the previous examples, under the central model, the Tsallis CDs  prove to be good competitors to the likelihood based CDs, while, under the contaminated model, the robust CDs present a better and robust performance. Figure~\ref{figsim333} (second column) reports the boxplots of the median of the likelihood and Tsallis based CDs, both under the central model and under a contaminated model. Also in this example, under the central model, the two estimators present a similar behaviour, while  only the Tsallis estimator present a robust performance with respect to contamination.

\begin{figure}
\begin{center}
\includegraphics[height=3cm,width=16cm]{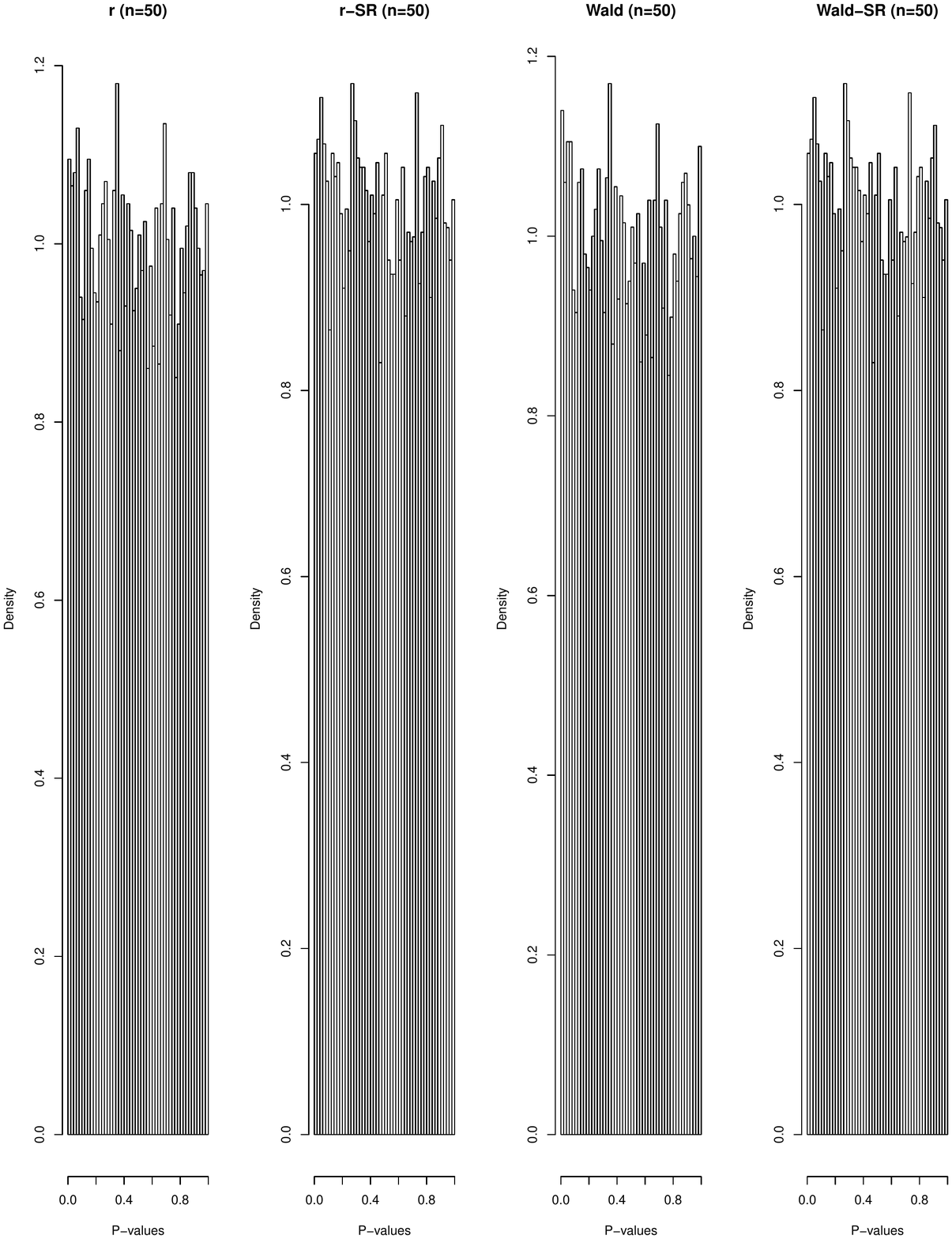} \\
\includegraphics[height=3cm,width=16cm]{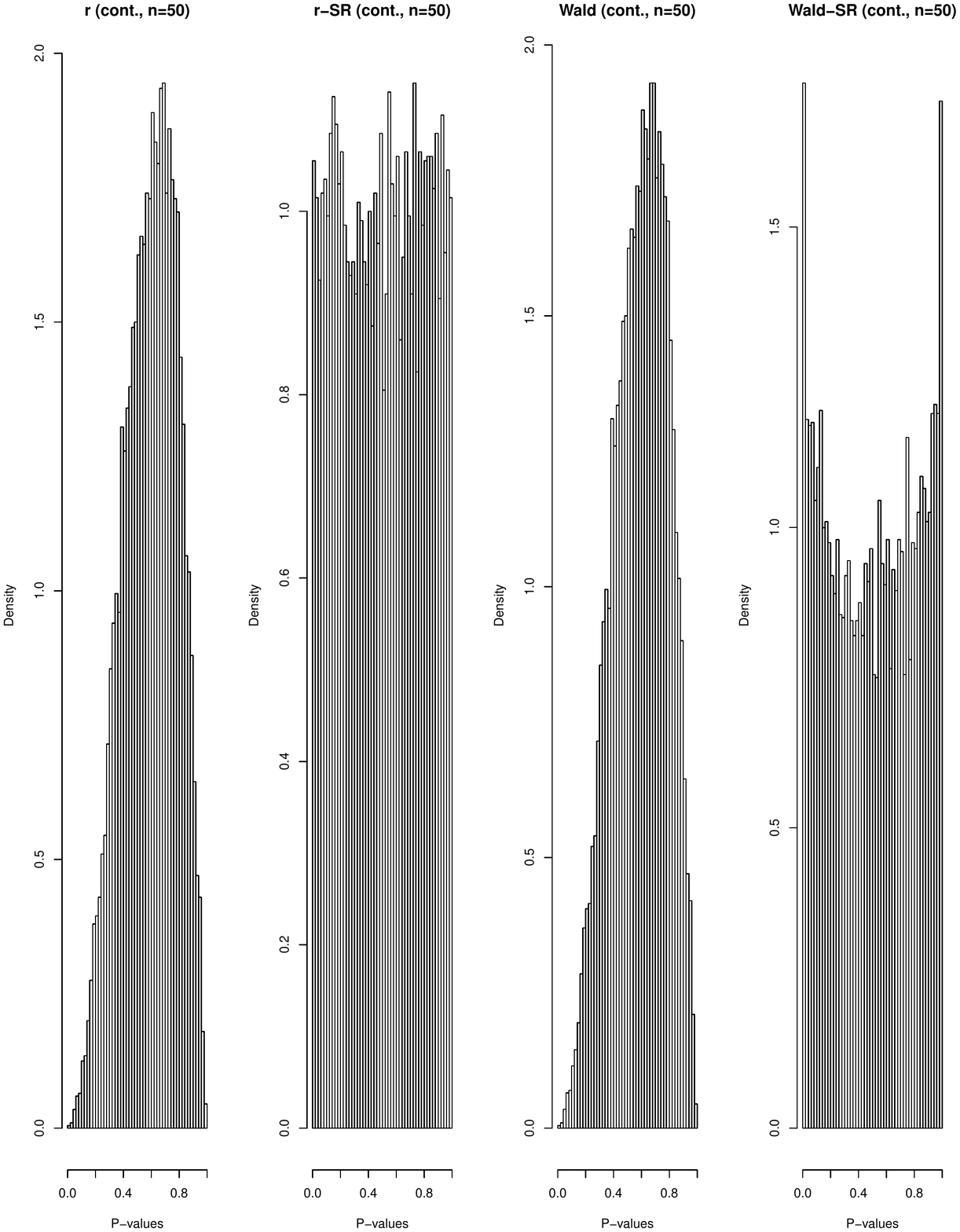}
\vspace{-0.3cm}
\caption{{\small Regression model. Empirical distribution of $p$-values for $H_0 : \psi = 0$ against $H_1 : \psi < 0$. Figures on the first row are based on non contaminated data, those of the second row are based on contaminated data.}}
\label{figsim555}
\end{center}
\end{figure}

Finally, Figure ~\ref{figsim555}report the uniform quantile-quantile plots of the p-values from the four CDs when testing $H_0: \psi=\psi_0$ against $H_1: \psi<\psi_0$, under the central model and under a contaminated model. We note that, under the central model all the CDs present a reasonable performance, while, under the contaminated model, only the robust CD presents a good performance. 

\vspace{0.2cm}

\begin{figure}
\begin{center}
\includegraphics[height=6cm,width=8cm]{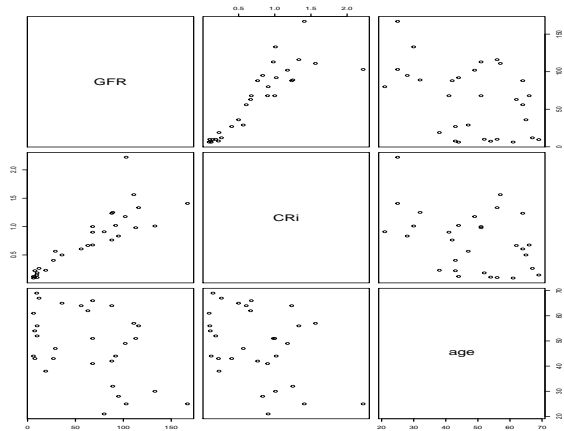}
\vspace{-0.3cm}
\caption{{\small Scatterplot diagram of GFR data.}}
\label{fig888}
\end{center}
\end{figure}

\noindent {\bf Case study.} The GFR dataset contains measurements of the glomerular filtration rate ($GFR$) and serum creatinine ($CR$) on $n = 30$ subjects. The $GFR$ is the volume of fluid filtered from the renal glomerular capillaries into the Bowmans capsule per unit of time (typically in millilitres per minute) and, clinically, it is often used to determine renal function. Its estimation is of clinical importance and several techniques are used for that purpose. One of them is based on $CR$, an endogenous molecule, synthesized in the body, which is freely filtered by the glomerulus (but also secreted by the renal tubules in very small amounts). Several models have been proposed in the literature to explain $GFR$ as a function of $CR$. Here, following Heritier {\em et al.} (2009), we consider a model for $GFR$ based on $CR^{-1}$ and $AGE$, i.e.  $GFR = \beta_1 + \beta_2 CR^{-1}  + \beta_3 AGE + \varepsilon$. The data are illustrated in Fig. \ref{fig888}: note that there are some observations which look like outliers.

\begin{figure}
\begin{center}
\includegraphics[height=6cm,width=8cm]{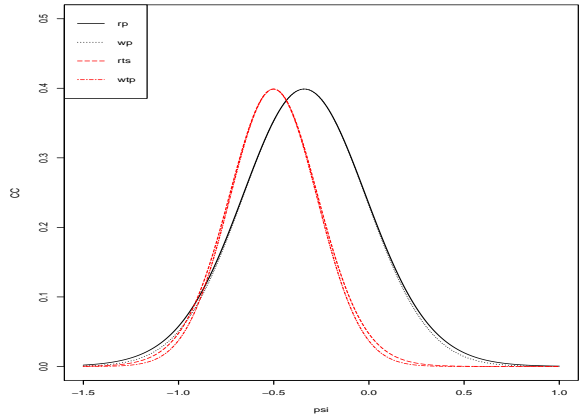}
\vspace{-0.3cm}
\caption{{\small CDs based on $r_p(\psi)$ and $w_p(\psi)$ and on $r_{Sp}(\psi)$ and $w_{Sp}(\psi)$ for $\psi=\beta_3$.}}
\label{fig999}
\end{center}
\end{figure}

Figure \ref{fig999} gives the CCs based on $r_p(\psi)$ and $w_p(\psi)$ and on $r_{Sp}(\psi)$ and $w_{Sp}(\psi)$ for $\psi=\beta_3$, i.e.\ the parameter of $AGE$. It can be noted that the robust CCs are quite different, giving different inferential conclusions about the effect on $AGE$ on $GFR$.

Table \label{tgfr} gives the p-values for $H_0 : \psi = 0 $ against $H_0: \psi \neq 0$, the measure of evidence for $\psi<0$ and the median of the CCs, based on $r_p(\psi)$ and on $r_{Sp}(\psi)$. Note that the CDs based on $r_p(\psi)$ and on $r_{Sp}(\psi)$ give quite different conclusions about the effect of $AGE$ on $GFR$.

\begin{table}
\begin{center}
\begin{tabular}{|c|c|c|c|} \hline
 CD                              &  $CC(0)$ &  p-value $H_0 : \psi = 0 $ & median\\ \hline
$\Phi(r_p(\psi))$          &  0.85        &   0.298                              & -0.33\\
$\Phi(r_{Sp}(\psi))$     &  0.98        &   0.039                              & -0.50\\ \hline
\end{tabular}
\vspace{0cm}
\caption{{\small CDs summaries based on $r_p(\psi)$ and on $r_{Sp}(\psi)$ for the GFR data and $\psi=\beta_3$.}}
\label{tgfr}
\end{center}
\end{table}

%%%%%%%%%%%%%%%%%%%%%%%%%%%%%%%%%%%%%%%%%
%%%%%%%%%%%%%%%%%%%%%%%%%%%%%%%%%%%%%%%%%

\section{Discussion}

In practical applications, CDs are more informative than a simpler confidence interval or a $p$-value, since they describe the complete  distribution estimator for the parameter of interest, as the posterior distribution for bayesians.
We would like to stress that, under appropriate smoothness conditions, the Tsallis scoring rule can be applied to any statistical model $f(y;\theta)$ and delivers an associated $M$-estimator. While this may lead to a loss of efficiency in comparison with full likelihood methods, it can exhibit improved robustness or computational advantages.  Moreover, under smoothness conditions, any proper scoring rule can be used to derive a CD, using the  first-order approximations of SR pivotal quantities. 

Higher-order asymptotic expansions for scoring rules have been recently discussed by Mameli and Ventura (2015) and Mameli {\em et al.}\ (2017). These higher-order asymptotic expansions to the distribution of the scoring rule estimator, of the scoring rule ratio test statistic and of the signed scoring rule root statistic for a scalar parameter allow to derive higher-order pivotal quantities, which improves the first-order approximations.  The use of these higher-order expansions could be investigated to derive CDs when dealing with small sample sizes.

Finally, in this paper only the median of the CDs has been considered. We are planning to investigate also other point estimators derived from the CDs, such as the mode, in particular in situation in which the CD exhibits a strong asymmetry.
%%%%%%%%%%%%%%%%%%%%%%%%%%%%%%%%%%%%%%%%%

%%%%%%%%%%%%%%%%%%%%%%%%%%%%%%%%%%%%%%%%%
\subsubsection*{\bf Founding}{\small This research work was partially supported by the University of Padova (\texttt{BIRD197903}) and by MIUR (PRIN 2015, grant \texttt{2015EASZFS\_003}).}

%%%%%%%%%%%%%%%%%%%%%%%%%%%%%%%%%%%%%%%%%

%\section*{References}


\begin{thebibliography}{99}
\bibitem{1}
Basu A, Harris IR, Hjort NL, Jones MC. Robust and efficient estimation by minimising a density power divergence. Biometrika, 1998; 85: 549--559.
\bibitem{2}
Basu A, Mandal A, Martin N, Pardo L.  Generalized Wald-type tests based on minimum density power divergence estimators. Statistics 2016;  50: 1--26.
\bibitem{3}
Brier GW (1950). Verification of forecasts expressed in terms of probability. Mon. Weather Rev., 1950; 78: 1--3.
\bibitem{4}
Dawid AP. Probability forecasting. In: {\em Encyclopedia of Statistical Sciences} (S. Kotz, N. L. Johnson, and C. B. Read eds.) 1986; 210--218.
\bibitem{5}
Dawid AP. Musio M. Theory and Applications of Proper Scoring Rules. Metron, 2014;  72: 169--183.
\bibitem{6}
Dawid AP, Musio M.  Bayesian model selection based on proper scoring rules (with discussion). Bayesian Analysis, 2015; 10: 479--521. 
\bibitem{7}
Dawid AP, Musio M, Ventura L. Minimum scoring rule inference. Scand.\ J.\ 
Statist., 2016; 43: 123--138.
\bibitem{8}
Farcomeni A, Ventura L (2012). An overview of robust methods in medical research. Stat.\ Meth.\ Med.\ Res., 2012; 21: 111--133.
\bibitem{9}
Field CA, Ronchetti E. {\em Small Sample Asymptotics}. IMS Monograph Series, Hayward (CA); 1991.
\bibitem{10}
Ghosh M, Basu A. Robust estimation for independent non-homogeneous observations using density power divergence with applications to linear regression. Electr.\ J.\ Statist., 2013;  7: 2420--2456.
\bibitem{11}
Ghosh M, Basu A. Robust Bayes estimation using the density power divergence. Ann.\ Inst.\ Stat.\ Math., 2016; 68: 413--437.
\bibitem{12}
Ghosh A, Martin N, Basu A, Pardo L. A new class of robust two-sample Wald-type
tests.  Int.\ J.\ Biostat., 2019; 20170023.
\bibitem{13}
Girardi P, Greco L,  Mameli V, Musio M, Racugno W,  Ruli E, Ventura L. Robust inference for nonlinear regression models from the Tsallis score: application to Covid-19 contagion in Italy. Stat
, 2020; 9:e309.
\bibitem{14}
Giummol\'e F, Mameli V, Ruli E, Ventura L. Objective Bayesian inference with proper scoring rules. Test, 2019;  28: 728--755.
\bibitem{15}
Good IJ. Rational decisions. J.\ Roy.\ Statist.\ Soc. B, 1952; 14: 107--114.
\bibitem{16}
Heritier S, Cantoni E, Copt S, Victoria-Feser MP ). {\em Robust Methods in Biostatistics}. Wiley; 2009.
\bibitem{17}
Heritier S, Ronchetti EM. Robust bounded-influence tests in general parametric models.   J.\ Americ. \ Statist. \ Assoc., 1994; 89: 897--904.
\bibitem{18}
Huber PJ, Ronchetti EM. {\em Robust Statistics}. Wiley, New York; 2009.
\bibitem{19}
Hyv\"arinen A. Estimation of non-normalized statistical models by score matching. Journal of Machine Learning Research, 2005;  6: 695--709.
\bibitem{20}
Hjort NL, Schweder T. Confidence distributions and related themes.J.\ Statist.\ Plan.\ Infer., 2018;  195: 1--13.
\bibitem{21}
Machete R. Contrasting probabilistic scoring rules. J.\ Statist.\ Plann.\ Inf., 2013; 143: 1781--1790.
\bibitem{22}
Mameli V, Musio M, Ventura L. Bootstrap adjustments of signed scoring rule root statistics. Comm.\ Statist.\ - Simul.\ Comput., 2018; 47:  4, 1204--1215.
\bibitem{23}
Mameli V, Ventura L. Higher-order asymptotics for scoring rules. J.\ Statist.\ Plann.\
 Inf., 2015; 165: 13--26.
\bibitem{24}  
 Mandas A. Congiu MG, Abete C, Dess\`i S, Manconi PE, Musio M, Columbu S., Racugno.  Cognitive decline and depressive symptoms in late-life are associated with statin use: evidence from a population-based study of Sardinian old people living in their own home. Neur.\ Res., 2014;  36: 3, 247--254. 
\bibitem{25}
Nielsen F, Nock R. A closed-form expression for the Sharma-Mittal entropy of exponential families. J. Phys. A: Math. Theor., 2012; 45: 032003.
\bibitem{26}
Pak RJ. The minimum density power divergence estimation for the lognormal density. 
Comm.\ Stat.\ - Theory and Methods, 2014;  43: 4582--4588.
\bibitem{27}
Pauli F, Racugno W, Ventura L. Bayesian composite marginal likelihoods. Statistica Sinica, 2011; 21: 149--164.
\bibitem{28}
Ronchetti E, Ventura L. Between stability and higher-order asymptotics. Stat.\ and Comput., 2001;  11: 67--73.
\bibitem{29}
Ruli E, Ventura L. Can Bayesian, confidence distribution and frequentist inference agree?.  Statistical Methods \& Applications, 2020; DOI s10260-020-00520-y.
\bibitem{30}
Schweder T, Hjort NL . {\em Confidence, Likelihood, Probability: Statistical Inference with Confidence Distributions}. Cambridge University Press; 2016.
\bibitem{31}
Tsallis C. Possible generalization of Boltzmann-Gibbs statistics. J.\ Statist.\ Physics, 1988; 52: 479--487.
\bibitem{32}
Varin C, Reid N, Firth D. An overview of composite likelihood methods. Statist.\ Sinica, 2011;  21:
5--42.
\bibitem{33}
Ventura L, Racugno W. Recent advances on Bayesian inference for $P(X < Y)$, Bayesian Analysis, 2011; 6: 411--428.
\bibitem{34}
Xie M, Singh K. Confidence distribution, the frequentist distribution estimator of a parameter: a review. Int.\  Statist.\ Rev., 2013; 81: 3--39.
\end{thebibliography}
\end{document}